\documentclass[aps,
	prd,
	superscriptaddress,
	preprintnumbers, 
	amssymb,
	amsmath,
	notitlepage,
	longbibliography,
	nofootinbib]{revtex4-1}
\usepackage[colorlinks=true, 
	linkcolor=blue, 
	citecolor=blue, 
	urlcolor=blue, 
	linktocpage=true]{hyperref}
\usepackage[utf8]{inputenc}
\usepackage[english]{babel}
\usepackage{stix}
\usepackage[final]{microtype}
\usepackage{xcolor}
\usepackage{graphicx}

\newcommand{\lp}{\ensuremath{\left(}}
\newcommand{\rp}{\ensuremath{\right)}}
\newcommand{\lb}{\ensuremath{\left\lbrace}}
\newcommand{\rb}{\ensuremath{\right\rbrace}}
\newcommand*{\s}{\nobreak\hspace{.08em plus .04em}}
\newcommand*{\dd}{\ensuremath{\mathrm{d}}}
\newcommand*{\e}{\ensuremath{\mathrm{e}}}
\newcommand*{\ii}{\ensuremath{\mathrm{i}}}
\newcommand*{\abs}[1]{\ensuremath{\left \lvert #1 \right\rvert}} 
\newcommand*{\pd}{\ensuremath{\partial}}
\newcommand*{\la}{\ensuremath{\lambda}}
\newcommand*{\ep}{\ensuremath{\epsilon}}

\begin{document}

\title{Parametrized Path Approach to Vacuum Decay}
\date{\today} 

\author{Florent Michel}
\email{florent.michel.10@normalesup.org}
\affiliation{Centre for Particle Theory, Durham University,
South Road, Durham, DH1 3LE, United Kingdom}

\begin{abstract} 
	We develop a new real-time approach to vacuum decay based on a reduction to a finite number of degrees of freedom. 
	The dynamics is followed by solving a generalized Schrödinger equation. 
	We first apply this method to a real scalar field in Minkowski space and compare the decay rate with that obtained by the instanton approach. 
	The main difference is in the early-time dynamics, where the decay is faster due to the tail of the wave function. 
	We then apply it to a cold atom model recently proposed to simulate vacuum decay experimentally. 
	This approach will be extended to include gravity in a future work.
\end{abstract}

\maketitle

\section{Introduction}

First-order phase transitions are ubiquitous in Nature, from water vapour turning into clouds to possible QCD-Electroweak transitions in the early Universe~\cite{PhysRevLett.119.141301}. 
Yet, the absence of a critical point and dependence on the microscopic degrees of freedom make their dynamics challenging to understand. 
These difficulties are particularly prominent in the case of \textit{quantum} first-order transitions, triggered by quantum fluctuations and involving deeply nonperturbative effects. 
They are also referred to as vacuum decay, \textit{i.e.}, the decay from a “false”, metastable vacuum (a local minimum of the relevant potential) to the “true” vacuum, \textit{i.e.}, the ground state of the theory.

In relativistic theories, vacuum decay generally occurs in two steps~\cite{Coleman:1977py}. 
Starting from the false vacuum phase, bubbles of true vacuum first nucleate through quantum and/or statistical fluctuations. 
They then expand, at a speed close to the speed of light, until they merge and fill the whole available space with true vacuum.
One important quantity to understand the dynamics is thus the rate of formation of such bubbles, which is also the decay rate of the false vacuum phase. 
The standard approach to computing the decay rate was laid out in a series of articles by Sydney Coleman and collaborators in the '70s \cite{Coleman:1977py, Callan:1977pt, Coleman:1980aw}. 
At its core is a saddle-point approximation of the path integral, expected to be valid provided the action of relevant solutions of the field equations is large before the reduced Planck constant. 
In Ref~\cite{Coleman:1977py}, Coleman motivates that the Wentzel-Kramers-Brillouin expansion used in non-relativistic quantum mechanics to compute the time a particle takes to tunnel through a potential barrier can be extended to the quantum theory of a relativistic scalar field in flat space-time. 
Like in the non-relativistic problem, the calculation involves finding classical solutions in imaginary time, called \textit{instantons} in this context. 
These solutions are stationary points of the Euclidean action, obtained from the standard action by replacing the (real) time coordinate by an imaginary one.
A saddle-point approximation of the path integral for the transition between the false and true vacua then gives the decay rate $\Gamma$ as a function of the difference $\Delta S_E$ in Euclidean action between the instanton and false vacuum: 
\begin{equation}
	\Gamma = A \s \e^{- \Delta S_E / \hbar},
\end{equation}
where $A$ is a positive number with dimension inverse of time such that, if $t_0$ and $l_0$ are typical time- and length-scales for the problem%
~\footnote{If $V_0$ is a typical potential scale, one may choose $c \s t_0 \sim l_0 \sim \lp V_0 / (c \s \hbar) \rp^{-1/(d+1)}$, where $c$ is the celerity of light.}
and at finite volume $\mathcal{V}$, 
\begin{equation}
	\abs{\ln \lp \frac{A \s t_0 \s l_0^d}{\mathcal{V}} \rp} \ll \frac{\Delta S_E}{\hbar}
\end{equation}
in the semi-classical limit $\Delta S_E \gg \hbar$, where $d$ is the number of space dimensions.
In Ref~\cite{Callan:1977pt}, Coleman and Callan compute the first quantum corrections, \textit{i.e.}, the coefficient $A$ to leading order in $\hbar / \Delta S_E$. 
This coefficient can be written as a product of two factors: one coming from translation invariance and equal to $\lp \Delta S_E / (2 \s \pi \s \hbar) \rp^{n/2}$, where $n$ is the number of directions in which the model is invariant, and one involving the determinants of the fluctuations operators defined over the instanton%
\footnote{after excluding the zero modes associated with translation invariance, which are taken into account by the first factor }
and false vacuum. 
The analysis of~\cite{Coleman:1977py} was then extended by Coleman and De Luccia in~\cite{Coleman:1980aw} to include gravitational effects. 

This topic has gained prominence thanks to the calculation of the effective potential of the Higgs boson reported in~\cite{Degrassi:2012ry}, which suggests that the current Higgs vacuum may be only a metastable state, \textit{i.e.}, a false vacuum. 
The true vacuum would lie at a larger value of the Higgs mass, possibly close to the Planck mass, raising the question of why, after billions of years of cosmic evolution, the visible universe has not yet decayed to the true vacuum. 
Besides anthropic arguments or extensions of the standard model of particle physics, a natural answer until a few years ago was that the decay rate, computed in homogeneous space, is smaller than the inverse age of the Universe due to the large potential barrier separating the current Higgs vacuum from the region of lower potential. 
However, it was shown \cite{Heinemeyer:2013tqa, Gregory:2013hja, Burda:2015yfa, Gregory:2016xix, Burda:2016mou, Mukaida:2017bgd, Gregory:2018bdt, Cuspinera:2018woe, Moss:2018rvu, Oshita:2018ptr, 2019arXiv190901378O, Cuspinera:2019jwt} that inhomogeneities can dramatically change this result.
In particular, the presence of a single sufficiently small black hole could, assuming possible extensions of the standard model couple weakly to the Higgs boson and that the instanton approach remains valid in curved space-times, be enough to trigger the transition to the true vacuum. 
These results (and the fact that the Universe as we know it, by definition, still exists), can be used to put strong bounds on the number density of primordial black holes and possible extensions of the standard model.

The use of the instanton approach, however, suffers from some difficulties. 
First, to our knowledge, a derivation exists only in flat space-times; the extension to curved ones proceeds by analogies rather than first principles. 
A possibly related issue is the number of negative modes: while it was shown~\cite{Coleman:1987rm} that instantons in Minkowski space generally have exactly one negative mode, which is crucial in their interpretation~\cite{Callan:1977pt}, no such result exists in curved space-times. 
In fact, some instanton solutions were shown to sustain infinitely many negative modes (see for instance \cite{Lee:2014uza, Gregory:2018bdt}). 
The interpretation of these solutions in the context of vacuum decay is still an open question. 
Another issue is that the Wick rotation from real to imaginary times relies on a choice of time coordinate on which to perform the rotation, which is ambiguous in non-static space-times~\cite{Visser:2017atf}.
Finally, the instanton approach does not say anything about the dynamics of the field during the nucleation event~\cite{1989NuPhB.324..157B, 1991AmJPh..59..994B}.

Several methods have already been tried to solve these issues. 
One promising suggestion is that the metric, rather than the time coordinate, should be made complex~\cite{Visser:2017atf, BarberoG.:1995ud}. 
More recently, attempts have been made at developing a formalism using only real time coordinates in analogy with what can be done in non-relativistic Quantum Mechanics~\cite{Turok:2013dfa, PhysRevD.16.3507} or with numerical techniques used in cold atom like the truncated Wigner approximation~\cite{Braden:2017add, Braden:2018tky}. 
It was also proposed that instantons may be generalized using Picard-Lefschetz thimbles~\cite{Cherman:2014sba, Brown:2017wpl, Mou:2019tck, Mou:2019gyl, Ai:2019fri} or other functional techniques~\cite{Bitar:1977wy, Tye:2009rb, Darme:2019ubo, Espinosa:2019hbm}.

In this article, we will follow a different approach, with some overlap with the above ones but complementary in other ways. 
In spirit, it is similar to the one developed for non-Abelian gauge theories in~\cite{PhysRevD.17.486}, and could prove useful to study the large-order behaviour of perturbation theory, following arguments given in~\cite{RUBAKOV1995245}.~\footnote{I thank an anonymous referee for bringing these references to my attention.}
The main idea will be to perform a systematic reduction to a finite set of degrees of freedom, effectively reducing the field-theoretic problem to finite-dimensional Quantum Mechanics. 
Once the initial state is chosen, its evolution will be followed by solving a Schrödinger equation, and the decay rate extracted from its solutions.
From a more formal point of view, this amounts to finding a path or set of paths in field configuration space interpolating between the false and true vacua, which is optimal according to some measure, before quantizing parameters along it.
We shall refer to this procedure as the \textit{parametrized path approach} to vacuum decay.
In this article we will present the idea of the procedure, show how it works on the simplest case of a real scalar field in flat space as well as in an “analogue” cold atom model, and compare its results with those of the instanton approach. 

In both cases, we find two qualitatively distinct behaviours. 
At late times, the decay rate tends towards a constant close to the result from the instanton approach, and apparently independent of the initial state. 
At early times, however, the decay rate takes much larger values due to the tail of the initial wave function extending beyond the potential barrier. 
This is related to the absence of exact stationary state centred on the false vacuum. 
We conjecture this early dependence on the tail of the initial quantum state persists in a field-theoretic treatment of the problem.

Extensions to more realistic scenarios and to vacuum decay in the presence of gravity will be dealt with in follow-up publications.
Unless stated otherwise, we work in natural units where the speed of light in vacuum, reduced Planck constant, and Boltzmann constant are equal to $1$.

\section{Relativistic scalar field in flat space}
\label{sec:Rsf}

In this section we develop the formalism for the parametrized path approach to tunnelling in the simplest case of a real scalar field in Minkowski space. 
For the sake of generality, we work in any number $d$ of space dimensions. 
The metric signature is $(+,-,-,\dots,-)$. 
In the following, for any natural integer $n$, $\mathcal{A}_n$ denotes the area of the Euclidean $n$-sphere with unit radius. 
(For instance, $\mathcal{A}_1 = 2 \s \pi$ and $\mathcal{A}_2 = 4 \s \pi$.) 
We set $\mathcal{A}_0 = 2$, so that all the expressions below remain valid for $d=1$.

\subsection{The model}

We consider a real scalar field $\phi$ in $(1+d)$-dimensional Minkowski space with potential $V$. 
The action is: 
\begin{equation}\label{eq:action}
	S(\phi) = \int \left[
		\frac{1}{2} \s \big( \pd^\mu \phi(x) \big) \big( \pd_\mu \phi(x) \big)
		- V \lp \phi(x) \rp
		\right] \dd x^{d+1},
\end{equation}
where Einstein's summation convention on repeated indices is used. 

For simplicity, we concentrate on solutions with an $\mathrm{O} (d-1)$ symmetry in one inertial frame.% 
\footnote{This assumption is justified by Ref~\cite{Coleman:1977th}, where it is shown that, under generic assumptions, the solution of the Euclidean field equation minimizing the action in flat space is spherically symmetric. 
Extension of the formalism to non-spherically symmetric configurations may be relevant in cases where the assumptions of this reference are not satisfied, but is left for a future work.}
We choose an inertial frame and a point $O$ stationary in this frame, and consider field configurations depending only of the time coordinate $t$ and distance $r$ from $O$. 
The action becomes: 
\begin{equation}\label{eq:action_rt}
	S(\phi) = \mathcal{A}_{d-1} \int_{t=-\infty}^{+\infty} \int_{r=0}^\infty \left[ 
		\frac{1}{2} \s \lp \pd_t \phi(t,r) \rp^2
		- \frac{1}{2} \s \lp \pd_r \phi(t,r) \rp^2
		- V \lp \phi(t,r) \rp
	\right] r^{d-1} \s \dd r \s \dd t . 
\end{equation}

We assume the potential $V\,$ has two minima: a global one at a value $\phi_T$ of $\phi$ and a local one at a value $\phi_F$. 
Homogeneous configurations with $\phi = \phi_T$ or $\phi = \phi_F$ are then saddle points of the action~\eqref{eq:action}, and thus solutions of the classical field equations. 
The first one is the true vacuum (TV) and the second one is the false vacuum (FV). 
We are interested in the phase transition from the later to the former through nucleation of a spherical bubble of TV in the FV phase. 
As mentioned in the introduction, this process can be described by finding saddle points of the action~\eqref{eq:action_rt} in imaginary and real times, describing respectively the quantum nucleation of the bubble and the ensuing classical evolution. 
In particular, defining the Euclidean time $\tau \equiv \ii \s t$ and Euclidean action%
\footnote{To simplify the notations, we use the same letter $\phi$ for the field before and after Wick rotation $t \to \tau$. 
The two cases are distinguished by the time coordinate, which will always be called $t$ before and $\tau$ after the Wick rotation.}
\begin{equation}\label{eq:action_rt_E}
	S_E(\phi) = \mathcal{A}_{d-1} \int_{\tau=-\infty}^{+\infty} \int_{r=0}^\infty \left[ 
		\frac{1}{2} \s \lp \pd_\tau \phi(\tau,r) \rp^2
		+ \frac{1}{2} \s \lp \pd_r \phi(\tau,r) \rp^2
		+ V \lp \phi(\tau,r) \rp
	\right] r^{d-1} \s \dd r \s \dd \tau,
\end{equation}
the decay rate $\Gamma$ of the false vacuum may be written as~\cite{Coleman:1977py,Callan:1977pt}
\begin{equation}\label{eq:Gamma_Col}
	\Gamma = A \exp \lp S_E(\phi_F) - S_E(\phi_B) \rp,
\end{equation}
where $A$ is a positive number and $\phi_B$ is the instanton solution of the field equation going to the FV as $\tau \to \pm \infty$ or $r \to \infty$, also called the “bounce”.%
\footnote{In this expression, it is assumed that the integrals defining $S_E(\phi)$ and $S_E(\phi_F)$ are convergent. 
In practice, this is generally the case if the zero of the potential $V$ is set at $\phi_F$.} 
The prefactor $A$ is proportional to the volume of a hypersurface of constant $t$, rendering the result formally infinite. 
Of course, the physically relevant quantity is the decay rate \textit{per unit volume}, which is finite.
We now describe a different way to approach this problem. 

\subsection{Parametrized path: generalities}
\label{sub:PP:gen}

We propose to estimate the decay rate of the false vacuum in the following way. 
We shall first choose a path in the space of field configurations interpolating between the FV and TV phases. 
Ideally, this path should be close to the one minimizing the action. 
A valley of the action functional may be obtained numerically using techniques close to those of~\cite{Darme:2019ubo,Espinosa:2019hbm}, which would provide, in some sense, an optimal path. 
However, here we shall instead use an \textit{Ansatz} with a simple analytical form. 
The reason for this choice is two-fold. 
First, having an analytical formula for the \textit{Ansatz} greatly simplifies the numerical calculations of Section~\ref{sec:num}. 
Second, it allows for comparison between results obtained with different Ansätze (as done in Appendices~\ref{app:asy} and~\ref{app:Rdw}), which can be used to test the robustness of the results. 
It would be interesting, but beyond the scope of the present work, to see more precisely the links and differences between this approach and the functional techniques of the above references.  

Once the \textit{Ansatz} is chosen, we shall quantize one or several of its parameters. 
We will then write a Schrödinger equation for this finite set of degrees of freedom. 
Finally, we shall determine initial conditions and solve the Schrödinger equation numerically. 
The numerical resolution will be described in Section~\ref{sec:num}. 
For the initial conditions, a natural choice at zero temperature will be the ground state in a quadratic potential approximating the real one close to the FV. 
When working at finite temperature, we shall solve the Schrödinger equation several times starting with ground or excited states of the quadratic potential, before averaging observables over all simulations with a Boltzmann weight. 
We now make these ideas more precise. 

For simplicity, in this work we will quantize only one parameter $R$, whose absolute value is interpreted as the radius of the bubble. 
The generalisation to several degrees of freedom will be done in a future study. 
The parametrized path $\lp \phi_R \rp_{R \in \mathbb{R}}$ is a set of functions $\phi_R$ from $\mathbb{R}_+$ to $\mathbb{R}$ labelled by a real parameter $R$ satisfying the six properties:% 
~\footnote{These properties ensure the quantities defined below are finite, that the path interpolates between the FV and TV as $R$ goes from $0$ to $\pm \infty$, and that the functions $K$ and $U$ defined by Equations~\eqref{eq:K} and~\eqref{eq:U} satisfy $U(0) = 0$, $U'(0) = 0$, $U''(0) > 0$, $K(0) \neq 0$, and $K'(0) = 0$. 
They can be relaxed, as will be done in Section~\ref{sec:num}, provided these properties are satisfied.}
\begin{enumerate}
	\item The function $(R, r) \mapsto \phi_R(r)$ is twice differentiable. 
	\item For each $r \geq 0$, $\phi_0(r) = \phi_F$. 
	\item For each $r \geq 0$, $\phi_R(r) \to \phi_T$ as $R \to \pm \infty$. 
	\item The function $\pd_R \phi_R$ evaluated at $R = 0$ is nonvanishing in at least a finite interval.
	\item For each $r \geq 0$, $\pd_R^2 \phi_R(r) = 0$ for $R = 0$.
	\item For each $R$, $\phi_R(r)$ goes to $\phi_F$ as $r \to \infty$, its first derivatives with respect to $r$ and $R$ as well as $\pd_r \pd_R \phi_R(r)$ go to $0$, and the convergence is fast enough for $\lp \pd_r \phi_R \rp^2$, $\lp \pd_R \phi_R \rp^2$, $\lp \pd_r \pd_R \phi_R(r) \rp^2$, and $V \lp \phi_R \rp - V \lp \phi_F \rp$ to be integrable for the integration measure $r^{d-1} \s \dd r$.
\end{enumerate}
We now see $R$ as a dynamical variable depending on $t$, \textit{i.e.}, we consider field configurations of the form
\begin{equation}
	\phi (t,r) = \phi_{R(t)}(r)
\end{equation}
for some real-valued, twice differentiable function $R$. 
Let us define the two functions
\begin{equation}\label{eq:K}
	K(R) = \mathcal{A}_{d-1} \int_0^\infty
		\lp \pd_R \phi_R(r) \rp^2
	r^{d-1} \s \dd r
\end{equation}
and 
\begin{equation}\label{eq:U}
	U(R) = \mathcal{A}_{d-1} \int_0^\infty \lp
		\frac{1}{2} \s \lp \pd_r \phi_R(r) \rp^2
		+ V \lp \phi_R(r) \rp
		- V \lp \phi_F \rp
	\rp r^{d-1} \s \dd r. 
\end{equation}
The action~\eqref{eq:action_rt} becomes, after subtracting the contribution from the false vacuum: 
\begin{equation}\label{eq:Sa}
	S(\phi) - S(\phi_F) = \int_{-\infty}^{+\infty} \left[ 
		\frac{K(R(t))}{2} \s \lp R'(t) \rp^2
		- U(R(t))
	\right] \dd t. 
\end{equation}
Differentiation with respect to $R'$ gives the momentum $\Pi$ conjugate to $R$: 
\begin{equation}
	\Pi = K(R) \s R'.
\end{equation}
The classical Hamiltonian is thus: 
\begin{equation}
	H = \frac{\Pi^2}{2 \s K(R)} + U(R). 
\end{equation}

We quantize this system by promoting $R$ and $\Pi$ to hermitian operators $\hat{R}$ and $\hat{\Pi}$ satisfying the commutation relation: 
\begin{equation}
	\left[ \hat{R}, \hat{\Pi} \right] = \ii.
\end{equation}
There is, unfortunately, no unique way to extend the classical Hamiltonian to a hermitian operator. 
For instance, one can choose any Hamiltonian of the form
\begin{equation} \label{eq:Hamb}
	\hat{H} = \frac{1}{2} \hat{U}_1^\dagger \hat{\Pi} \s \lp K(\hat{R}) \rp^{-1} \s \hat{\Pi} \s \hat{U}_1
		+ \hat{U}_2^\dagger \s U(\hat{R}) \s \hat{U}_2.
\end{equation}
where $\hat{U}_1$ and $\hat{U}_2$ are two unitary operators. 
We refer to~\cite{Ali:2004ft} for a more in-depth discussion of operator ordering ambiguities. 
The correct Hamiltonian to consider should be derivable from the full quantum field theory by tracking the relations between the radius $R$ of the bubble and the creation and annihilation operators of field excitations, but this is beyond the scope of the present work which focuses on generic quantum effects rather than a precise description of a specific theory.  
Here, for simplicity we choose the Hamiltonian operator 
\begin{equation}
	\hat{H} = \frac{1}{2} \s \hat{\Pi} \s \lp K(\hat{R}) \rp^{-1} \hat{\Pi} + U(\hat{R}),
\end{equation}
corresponding to setting $\hat{U}_1 = \hat{U}_2 = 1$ in Equation~\eqref{eq:Hamb}.

Let $\psi$ be the wave function. 
Identifying $\hat{R}$ with multiplication by $R$ and $\hat{\Pi}$ with $- \ii \s \pd_R$ gives the Schrödinger equation: 
\begin{equation}\label{eq:Sc}
	\ii \s \pd_t \psi(t,R) = - \pd_R \lp \frac{\pd_R \psi(t,R)}{2 \s K(R)} \rp + U(R) \s \psi(t,R). 
\end{equation}
We have thus reduced the problem to that of a non-relativistic point particle with position $R$ and $R$-dependent mass $K$ moving in the potential $U$. 
We may now pause and reflect on the meaning of this reduction. 
Its validity will depend on the precise choice of \textit{Ansatz}: if the latter is too far from the actual configuration maximizing the decay rate (which should be close to the instanton in the regime where the instanton approach is valid, see~\cite{Coleman:1977py}), one can expect this procedure to overestimate the potential barrier between the FV and TV; if the \textit{Ansatz} is sensibly chosen, however, we may expect to obtain a good approximation of the full field-theoretic problem. 
Ultimately, validation will come from comparison of the decay rate with the one obtained by finding an instanton, for potentials in which the latter is expected to give an accurate estimate. 
The main merit of the present approach, in our opinion, is to separate the calculation into two distinct questions: the determination of the optimal path in field configuration space and the dynamical evolution. 
These can then be dealt with separately. 
In this work, we mostly focus on the dynamical aspects. 
The results reported in Section~\ref{sec:num} can in principle be improved by a more careful choice of \textit{Ansatz}, as mentioned above. 

Let us briefly discuss the behaviour of the effective potential $U$. 
From the above assumptions, we have $U(0) = U'(0) = 0$. 
Besides, since $\phi_F$ is a local minimum of $V$, a small perturbation around the false vacuum will always increase $U$; one can show that the variation does not vanish to second order. 
So, $U''(0) > 0$ and $U(R)$ can be approximated by a harmonic potential for $R \approx 0$. 
Besides, for large values of $\abs{R}$, $\phi_R(r)$ will typically be close to $\phi_F$ for $r \gg \abs{R}$ and to $\phi_T$ for $r \ll \abs{R}$. 
Assuming the width of the transition region and the derivative of the field do not grow too fast,%
\footnote{For instance, we may assume the former grows slower than linearly in $R$ and that $\abs{\pd_r \phi_R}$ remains smaller than a positive constant independent of $R$.}
then 
\begin{equation}\label{eq:asyR}
	U(R) \mathop{\sim}_{R \to \pm \infty} \frac{\mathcal{A}_{d-1}}{d} \s \lp V_T - V_F \rp \abs{R}^d.
\end{equation}
So, $U(R)$ has a local minimum at $R = 0$, at least one maximum, and goes to $-\infty$ as $\abs{R} \to \infty$. 
(An example is shown in the right panel of Figure~\ref{fig:UandK}.) 
Besides, under similar assumptions,% 
\footnote{For instance, one may assume $\pd_R \phi_R(r)$ is localized around $r=R$ and its square integral becomes independent of $R$ in the large-$R$ limit.}
$K(R)$ is typically proportional to $\abs{R}^{d-1}$ as $\abs{R} \to \pm \infty$. 
Equating the magnitudes of the kinetic and potential terms in the Schrödinger equation~\eqref{eq:Sc} thus gives a wavelength scaling like $\abs{R}^{(1 - 2 \s d) / 2}$ for $R \to \pm \infty$. 
Smaller wavelengths thus become prominent for large values of $\abs{R}$, and solutions require a finer and finer grid to be correctly described numerically. 
We will come back to this issue in Section~\ref{sub:Idr}.  

Let us now turn to the initial conditions. 
As mentioned in the paragraph above, $U(0) = U'(0) = 0$ while $U''(0) > 0$. 
When considering a wave function localized around $R = 0$, \textit{i.e.}, a state close to the false vacuum, we can thus approximate $U$ by a harmonic potential and write
\begin{equation}
	U(R) \approx \frac{U''(0)}{2} \s R^2. 
\end{equation}
Moreover, with the above assumptions, we have $K(0) > 0$ and $K'(0) = 0$. 
We may thus replace the Schrödinger equation~\eqref{eq:Sc} by
\begin{equation}
	\ii \s \pd_t \psi(t,R) \approx - \frac{\pd_R^2 \psi(t,R)}{2 \s K(0)} + \frac{U''(0)}{2} \s R^2 \s \psi(t,R). 
\end{equation}
The eigenstates are those of a point particle with mass $K(0)$ in a harmonic potential. 
For any natural integer $n$, the $n$th excited state ($n=0$ corresponding to the ground state) is
\begin{equation}\label{eq:psin}
	\psi_n(R) = \frac{1}{\sqrt{2^n \s n!}} \s \lp \frac{\sqrt{K(0) \s U''(0)}}{\pi} \rp^{1/4} \s \e^{- \sqrt{ K(0) \s U''(0)} \s R^2 / 2} \s H_n \lp \lp K(0) \s U''(0) \rp^{1/4} \s R \rp,
\end{equation}
where $H_n$ denotes the $n$th Hermite polynomial.
In the following, we shall work at zero and at finite temperatures. 
At zero temperature, we shall take $\psi_0$ as initial wave function. 
At finite temperature $T > 0$, we will consider a mixed state where the probability $\mathcal{P}_n$ of being in the pure state $\raisebox{-0.03ex}{\scalebox{1}[1.05]{$\lvert$}} \psi_n \rangle$ with wave function $\psi_n$ and energy $\mu_n$ is proportional to $\e^{- \mu_n / T}$. 
Imposing $\sum_{n=0}^\infty \mathcal{P}_n = 1$ gives: 
\begin{equation}
	\mathcal{P}_n = \e^{-n \s \sqrt{U''(0) / K(0)} / T} \s \lp 1 - \e^{\sqrt{U''(0) / K(0)} / T} \rp. 
\end{equation}
The initial state has the density matrix
\begin{equation}
	\hat{\rho} = \sum_{n=0}^\infty \mathcal{P}_n \s \raisebox{-0.03ex}{\scalebox{1}[1.05]{$\lvert$}} \psi_n \rangle \langle \psi_n \raisebox{-0.03ex}{\scalebox{1}[1.05]{$\rvert$}}.
\end{equation}
In the numerical resolution, we truncate the sum to a finite number of excited states.
This choice of initial conditions should be relevant provided the typical width $\lp K(0) \s U''(0) \rp^{-1/4}$ of the ground state is much smaller than the first values of $R$ for which $U(R)$ differs significantly from the harmonic approximation, as the wave functions $\psi_n$ then correspond to quasi-stationary states. 
It is at the moment unclear to us how they should be generalized when it is not the case. 

\subsection{Comparison with the instanton approach} 

We want to compare results for the decay rate to those obtained using an instanton approach, \textit{i.e.}, by computing the action of a Euclidean solution. 
There are two ways to perform this calculation. 
The first one is to solve the field equation from the action~\eqref{eq:action_rt_E} in Euclidean time. 
We look for $\mathrm{O}(d+1)$ symmetric solutions. 
Up to a change of origin, one can thus assume $\phi$ depends only of $\rho \equiv \sqrt{\tau^2 + r^2}$, where $\tau = \ii \s t$ is the imaginary time. 
The Euclidean action $S_E$ is then
\begin{equation}\label{eq:SEi}
	S_E(\phi) = \mathcal{A}_d \int_0^\infty \left[ 
		\frac{1}{2} \s \lp \phi'(\rho) \rp^2
		+ V(\phi(\rho))
	\right] \rho^d \s \dd \rho
\end{equation}
and the Euler-Lagrange equation is
\begin{equation}\label{eq:inst_phi}
	\rho^{-d} \s \pd_\rho \lp \rho^d \s \phi'(\rho) \rp = V'(\phi(\rho)),
\end{equation}
with boundary conditions $\phi(\rho) \to \phi_F$ as $\rho \to \infty$ and $\phi'(0) = 0$. 

Another possibility is to start from the action~\eqref{eq:Sa}. 
Differentiating with respect to $R$ and replacing $t$ by $\tau = \ii \s t$ gives the ordinary differential equation: 
\begin{equation}\label{eq:R}
	\frac{\dd}{\dd \tau} \lp K(R(\tau)) \s R'(\tau) \rp = U'(R(\tau)),
\end{equation}
to be solved with the boundary condition $R(\tau) \to 0$ as $\tau \to \pm \infty$. 
An estimate for the Euclidean action of the instanton is then: 
\begin{equation}\label{eq:SEp}
	S_E \approx \int_{-\infty}^{+\infty} \left[ 
		\frac{K(R(\tau))}{2} \s \lp R'(\tau) \rp^2
		+ U(R(\tau))
	\right] \dd \tau .
\end{equation}
Equations~\eqref{eq:inst_phi} and~\eqref{eq:R} can be solved using a shooting method.

We thus have three ways to estimate the decay rate: by solving the Schrödinger equation~\eqref{eq:Sc}, by computing the instanton, or by solving the classical equation of motion after having chosen an \textit{Ansatz}. 
To obtain the numerical results reported below, we first compared the latter two as a check of the validity of the \textit{Ansatz}. 
We then compared their estimates with that obtained by solving the Schrödinger equation.

\section{Numerical results}
\label{sec:num}

In this section we show results obtained by solving numerically Equation~\eqref{eq:Sc} for a particular shape of the potential $V$. 
We focus on the case of a real scalar field in flat space, as discussed in Section~\ref{sec:Rsf}. 
Results for a cold atom analogue model will be shown in Section~\ref{sec:cond}. 
To make the comparison with the cold atom case easier, we work in two space dimensions. 
The method can be easily extended to different values of $d$. 
(The case $d=3$ is discussed in Appendix~\ref{app:3d}.)

\subsection{Choice of potential and \textit{Ansatz}}

\begin{figure}
	\centering
	\includegraphics[width=0.49\linewidth]{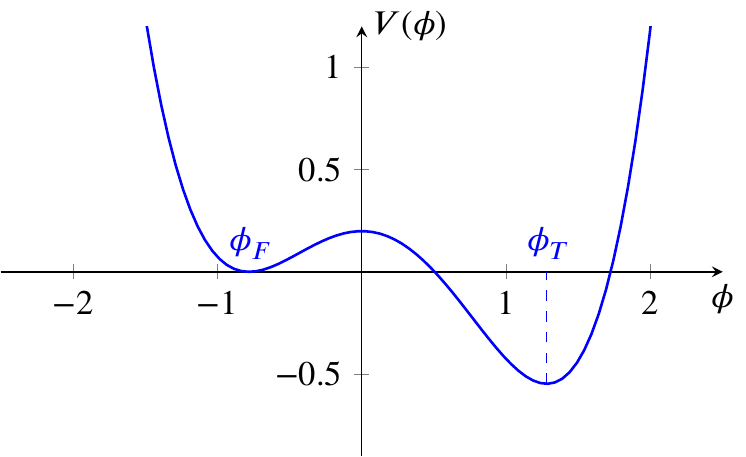}
	\caption{Sketch of the potential~\eqref{eq:pot} for $\la = 0.5$ and $\eta = 1.5$. 
		The values $\phi_F$ and $\phi_T$ are the false and true vacua, given by Equation~\eqref{eq:2:phiFT}.} 
	\label{fig:pot}
\end{figure}

We work with a potential of the form: 
\begin{equation}\label{eq:pot}
	V: \phi \mapsto \eta \s \left[ - \frac{\phi^2}{2} - \la \s \frac{\phi^3}{3} + \frac{\phi^4}{4} \right] - V_0,
\end{equation}
where $\eta$ and $\la$ are two positive numbers, and $V_0$ is a real number chosen to set the false vacuum energy to zero. 
This potential is sketched in Figure~\ref{fig:pot} for $\la = 0.5$ and $\eta = 1.5$.
It has three stationary points: a local maximum at $\phi = 0$ and two local minima at $\phi = \phi_F$ and $\phi = \phi_T$, where 
\begin{equation}\label{eq:2:phiFT}
	\phi_F = \frac{\la - \sqrt{\la^2  + 4}}{2}, \quad
	\phi_T = \frac{\la + \sqrt{\la^2  + 4}}{2}.
\end{equation}
Explicit calculation gives $V(\phi_F) > V(\phi_T)$. 
So, $\phi_F$ is the false vacuum and $\phi_T$ is the true vacuum. 
Roughly speaking, the parameter $\la$ controls the separation between the two minima, while $\eta$ sets the height of the potential barrier and depth of the minima.

\begin{figure}
	\centering
	\includegraphics[width=0.49\linewidth]{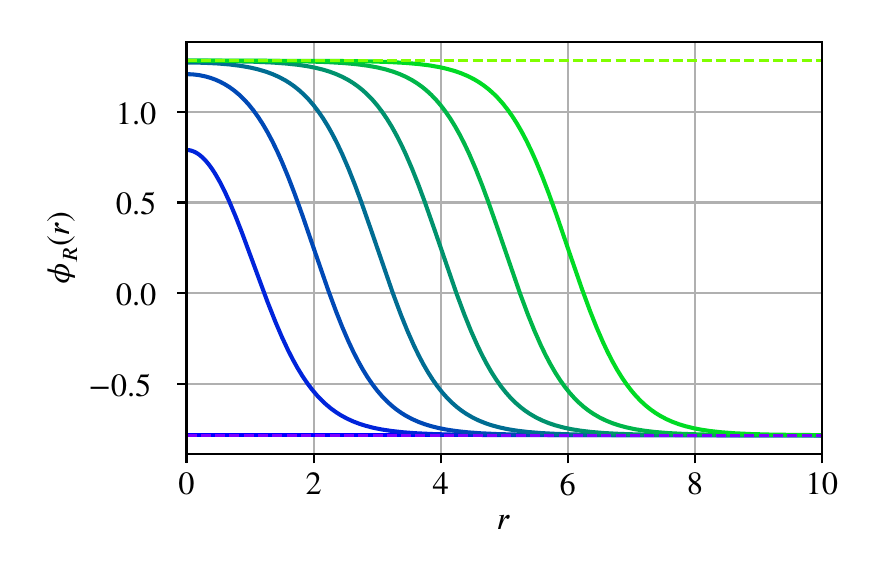}
	\caption{Illustration of the \textit{Ansatz}~\eqref{eq:10:anz3} for $\la = 1$ and $\sigma = 1$ for equally-spaced values of $R$ from $0$ (bluest curve) to $6$ (greenest curve). 
	The two dashed horizontal lines materialize $\phi = \phi_F$ (blue) and $\phi = \phi_T$ (green).}
	\label{fig:10:anz3}
\end{figure}

We choose the following \textit{Ansatz} for $\phi_R$, illustrated in Figure~\ref{fig:10:anz3}:
\begin{equation} \label{eq:10:anz3}
	\phi_R(r) = \phi_F + \frac{\phi_T - \phi_F}{2} \s \abs{ 
		\tanh \lp \frac{r + R}{\sigma} \rp
		- \tanh \lp \frac{r - R}{\sigma} \rp
	},
\end{equation}
where $\sigma > 0$ sets the typical width of the transition region between the two vacua. 
It satisfies all the above requirements, except that it is not twice differentiable in $R$ at $R=0$ because of the absolute value. 
Since $\lp \pd_R \phi_R \rp^2$ is differentiable in $R$ with a vanishing derivative at $R = 0$, one can show that all the properties derived in Section~\ref{sec:Rsf} still hold.
Results for a twice differentiable potential are shown in Appendix~\ref{app:asy}, showing small difference with those presented below. 
The case of an $R$-dependent width of the bubble wall is briefly discussed in Appendix~\ref{app:Rdw}.

Let us now discuss the choice of $\sigma$. 
As mentioned in Section~\ref{sec:Rsf}, the parameters of the \textit{Ansatz} should be chosen to be as close as possible to the path in field configuration space between the false and true vacua minimizing the Euclidean action. 
To achieve this, we define for each value of $\sigma$ the optimal Euclidean action $S_{E,\mathrm{opt}}(\sigma)$ equal to the right-hand side of Equation~\eqref{eq:SEp} for the function $R$ satisfying Equation~\eqref{eq:R} with boundary conditions $R(\tau) \to 0$ as $\tau \to \pm \infty$ and $R(0) \neq 0$. 
The value of $\sigma$ is then chosen by minimizing $S_{E,\mathrm{opt}}(\sigma)$. 
As a check of the validity of the \textit{Ansatz}, we verified that the corresponding Euclidean action is close to the right-hand side of Equation~\eqref{eq:SEi}.

\begin{figure}
	\centering
	\includegraphics[width=0.49\linewidth]{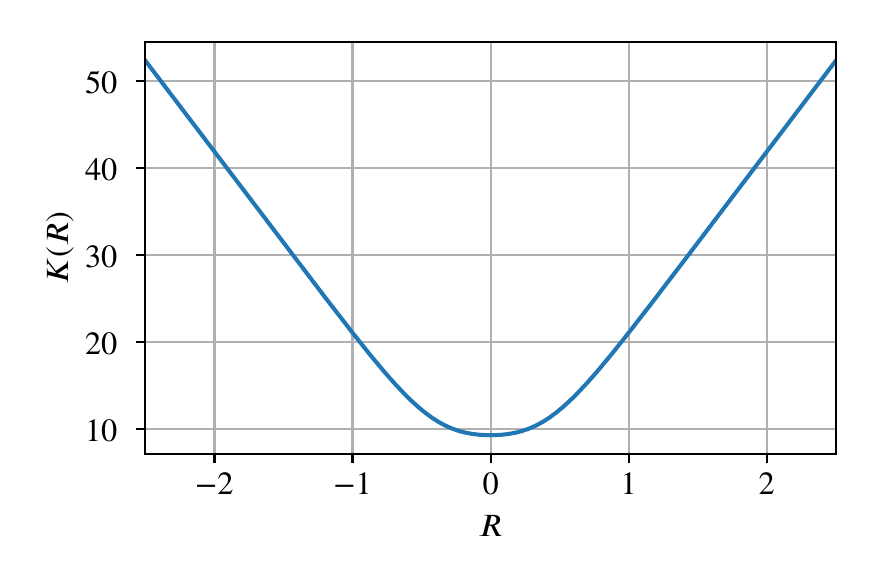}
	\includegraphics[width=0.49\linewidth]{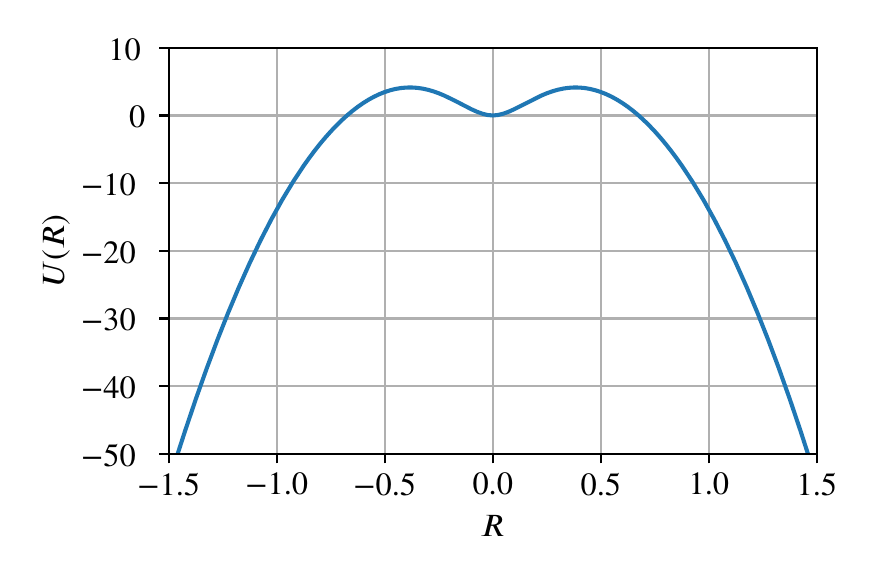}
	\caption{Plots of $K$ of Equation~\eqref{eq:K} (left panel) and $U$ of Equation~\eqref{eq:U} (right panel) as functions of $R$ for the potential~\eqref{eq:pot} with $\la=1$, $\eta=16$, and $\sigma = 0.5$.
	}
	\label{fig:UandK}
\end{figure}

Plots of the functions $U$ and $K$ are shown in Figure~\ref{fig:UandK}.
For all the parameters we tried, $K(R)$ has only one minimum at $R=0$ and goes monotonically to $\infty$ as $\abs{R}$ is increased, while $U(R)$ has three stationary points: a local minimum at $R = 0$ and two degenerate maxima at equal and opposite values of $R$. 
(Notice that the \textit{Ansatz} in invariant under $R \to -R$), and goes to $-\infty$ when $R \to \pm \infty$ as discussed in Section~\ref{sec:Rsf}.

\subsection{Instantaneous decay rate}
\label{sub:Idr}

Equation~\eqref{eq:Sc} was solved numerically using a finite difference method, implemented in Python 3.7 and C11. 
(The code is available on request.) 
As was mentioned in Section~\ref{sec:Rsf}, one difficulty is that the typical wavelength decreases as $\abs{R}$ increases away from the point where $U(R)$ reaches its maximum value, requiring an ever finer grid to be described accurately. 
In practice, this manifests itself as unphysical reflection of the perturbations from the large-$\abs{R}$ regions, where the wavelength becomes of the same order as the grid step, towards $R=0$. 
To circumvent this problem, we added a small dissipative term 
\begin{equation}
	- \ii \s \frac{\gamma}{2} \s \pd_R^4 \psi(t,R)
\end{equation}
to the right-hand side of Equation~\eqref{eq:Sc}, where $\gamma$ is a positive number. 
Its value was chosen to satisfy the two conditions: 
\begin{itemize}
	\item Modes with wavelengths close to the grid spacing should be damped within a few time steps.
	\item Modes with wavelengths of the order of a typical scale $R_0$ close to the width of the initial condition should suffer negligible damping during the whole evolution.
\end{itemize}
Denoting the space step by $\delta x$, the time step by $\delta t$, and the total duration of a simulation by $T$, these two conditions are equivalent to 
\begin{equation}
	\frac{\gamma}{2} \s \frac{16 \s \pi^4}{\delta x^4} \s \delta t \approx 1
	\qquad \text{and} \qquad
	\frac{\gamma}{2} \s \frac{16 \s \pi^4}{R_0^4} \s T \ll 1.
\end{equation}
The parameters are thus chosen so that 
\begin{equation}
	\frac{\delta x^4}{8 \s \pi^4 \s \delta t} \approx \gamma \ll \frac{R_0^4}{8 \s \pi^4 \s T}.
\end{equation}

Our main objective is to determine the instantaneous decay rate and its evolution in time. 
To this end, we first define $R_{U_{\mathrm{max}}}$ as the positive value of $R$ at which $U(R)$ reaches its maximum value. 
At a time $t$, the probability $P_F(t)$ that $R$ lies between the two maxima of $U$ is
\begin{equation} \label{eq:PF}
	P_F(t) = \int_{-R_{U_{\mathrm{max}}}}^{+R_{U_{\mathrm{max}}}} \abs{\psi(t,R)}^2 \s \dd R. 
\end{equation}
Intuitively, $P_F(t)$ may be seen as an estimate of the probability that the field remains close to the false vacuum at time $t$. 
There is some arbitrariness in choosing the boundaries of the integral; however, it should not affect the late-time behaviour of $P_F$ nor its qualitative profile.
We define the instantaneous decay rate $\Gamma$ as the opposite rate of change of $P_F$: 
\begin{equation}\label{eq:Gamma}
	\Gamma(t) = - \frac{1}{P_F(t)} \s \frac{\dd P_F(t)}{\dd t}. 
\end{equation}

\begin{figure}
	\centering
	\includegraphics[width=0.49\linewidth]{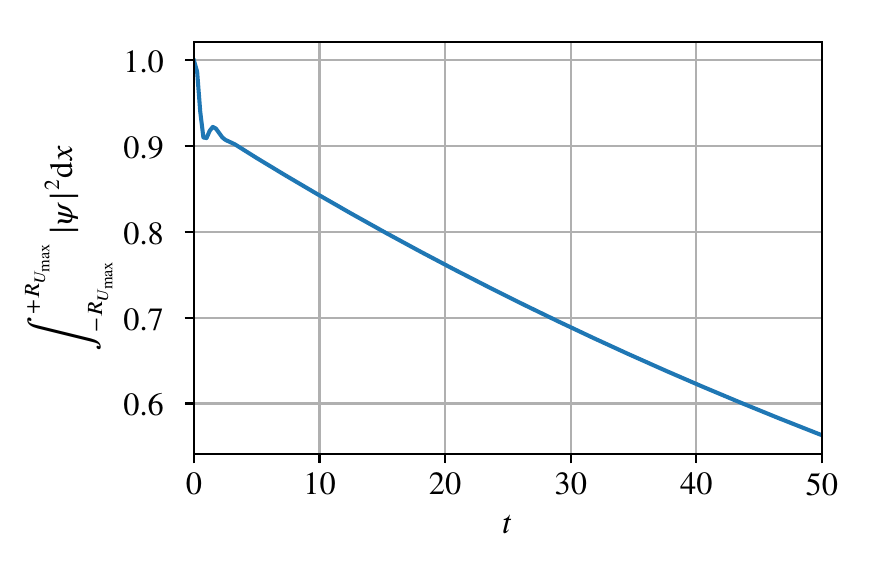}
	\includegraphics[width=0.49\linewidth]{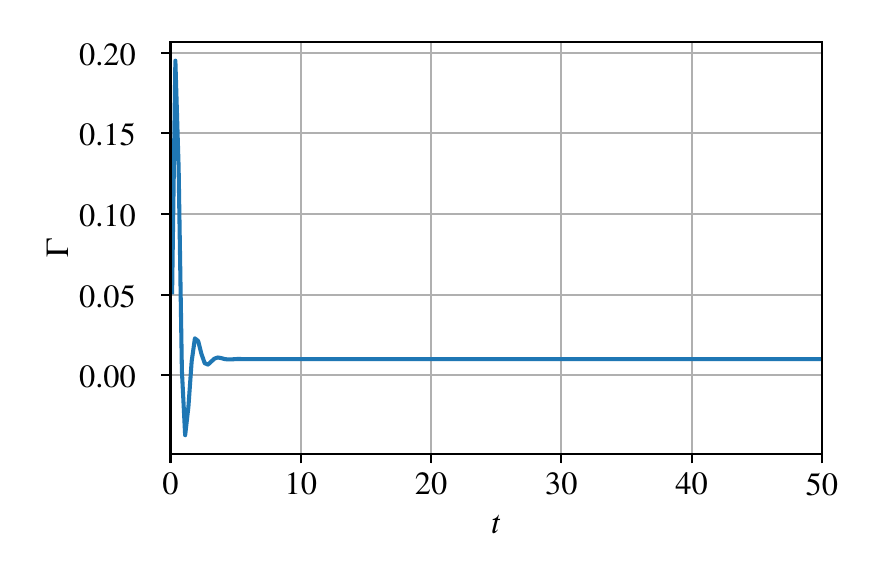}
	\caption{Results for the probability $P_F$ of remaining close to the false vacuum (left panel) and instantaneous decay rate (right panel) at zero temperature for $\la = 1$, $\eta = 16$, and $\gamma = 10^{-6}$.}
	\label{fig:10:pp3}
\end{figure}

Figure~\ref{fig:10:pp3} shows results from a typical simulation, done with $\la = 1$, $\eta = 16$, and $\gamma = 10^{-6}$. 
We work at zero temperature, with the initial condition set to $\psi_0$ of Equation~\eqref{eq:psin}. 
At early times, the instantaneous decay rate oscillates wildly, taking relatively large positive and even negative values. 
This seems to be due to the relatively wide initial wave function extending beyond the maxima of $U$. 
For $t > 5$, the initial time dependence subsides and the instantaneous decay rate becomes constant within numerical errors, with a value close to $10^{-2}$.

\begin{figure}
	\centering
	\includegraphics[width=0.49\linewidth]{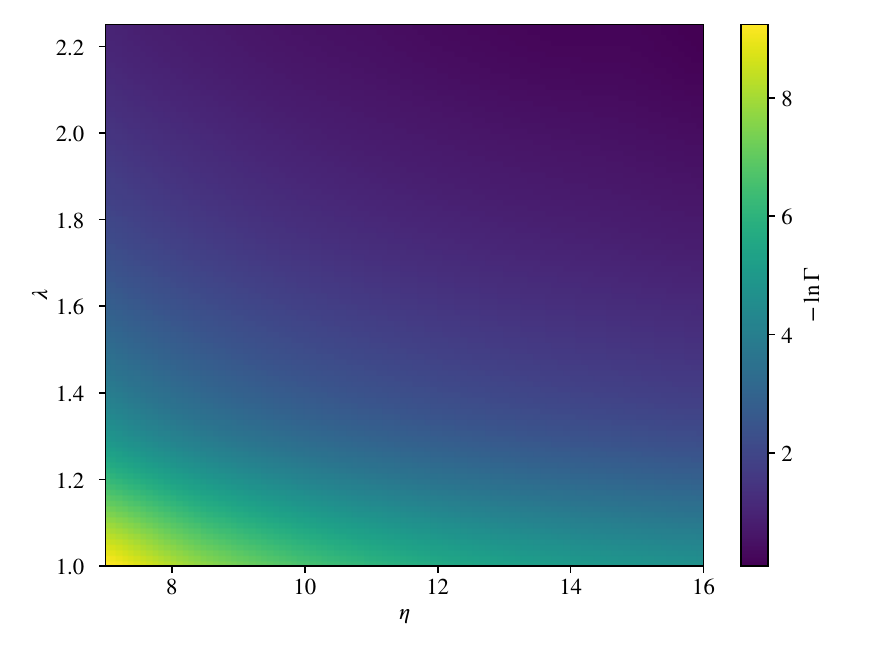}
	\includegraphics[width=0.49\linewidth]{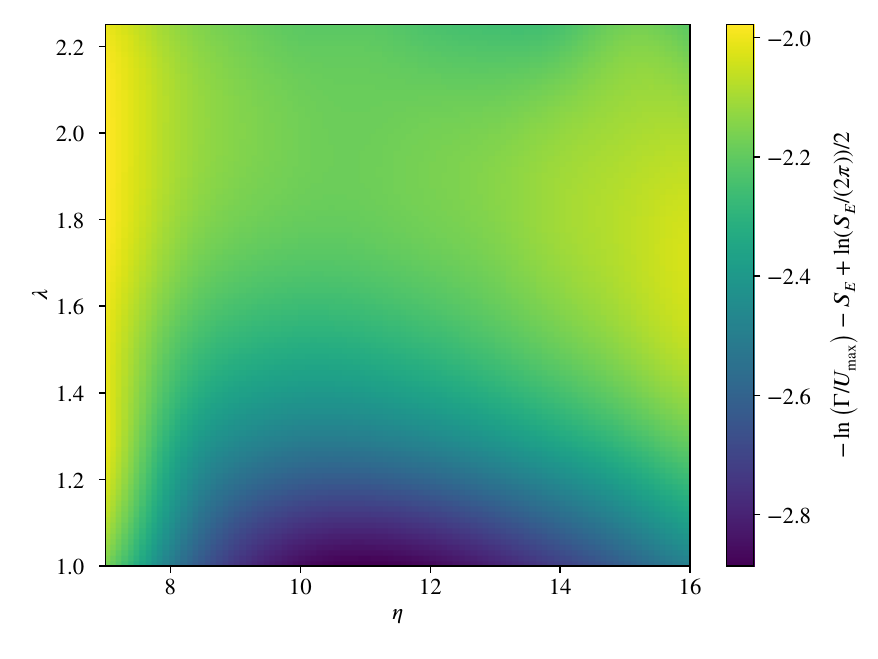}
	\caption{Plots of $-\ln \Gamma$ (left panel) and difference with the instanton action plus the contribution of the zero mode (right panel), where $\Gamma$ is evaluated at the latest time in each simulation, for $\la$ varying between $1$ and $2.3$, and $\eta$ varying between $7$ and $16$.
	The coefficient of the dissipative term is fixed at $\gamma = 5 \times 10^{-8}$.}
	\label{Fig:10:pp3b}
\end{figure}
Figure~\ref{Fig:10:pp3b} shows the late-time value of the decay rate for a range of values of $\eta$ and $\la$. 
In the left panel we show the value of $- \ln \Gamma$, which goes from a minimum value close to $0$ for large $\la$ to above $8$ for the smallest values of $\la$ and $\eta$. 
In the right panel, we adimensionalize $\Gamma$ by the maximum value $U_{\mathrm{max}}$ of the effective potential $U$ and subtract the Euclidean action $S_E$ of the instanton as well as the contribution from one zero mode (see~\cite{Callan:1977pt}).% 
\footnote{The motivation for including only one zero mode is that the \textit{Ansatz} fixes the position in space of the centre of the bubble, leaving only one translation invariance in the time direction.}
We notice that the range of variation is strongly reduced, with a minimum value close to $-2.8$ and a maximum one close to $-2.0$. 
It is also significantly smaller than that of the Euclidean action $S_E$, which varies between $2.0$ and $13.2$ for these values of the parameters.
This indicates that our approach is consistent with the instanton one, as it gives a decay rate with  similar order of magnitude and the same dependence in the two parameters. 
A more precise comparison would require to compute the full prefactor of the exponential in Equation~\eqref{eq:Gamma_Col}, which we do not undertake here. 
To our knowledge, a precise evaluation of this prefactor is at present unavailable.

\begin{figure}
	\centering
	\includegraphics[width=0.49\linewidth]{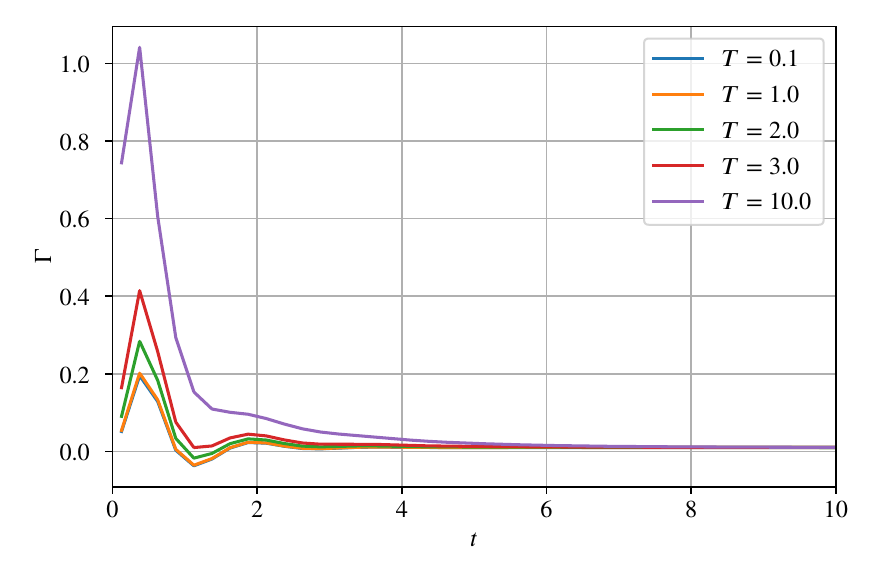}
	\caption{Instantaneous decay rate obtained for $\la = 1$, $\eta = 16$, and $\gamma = 10^{-6}$ for a temperature ranging from $0.1$ to $10$.}
	\label{fig:10:pp4}
\end{figure}
To end this section, we show in Figure~\ref{fig:10:pp4} results obtained at different finite temperatures $T$. 
For $T \leq 1$, the effect of the temperature on the decay seems to be very small. 
For $T > 1$, thermal effects increase the decay rate at early times, thus making the transition faster, but do not seem to affect its late-time limit. 
This is an important difference with the instanton approach, where instantons at finite temperature are expected to be periodic in Euclidean times (see for instance~\cite{Gross:1980br,Linde:1980tt,Khlebnikov:1991th,Chudnovsky:1992xrj}), giving a decay rate dependent on $T$. 
One possible resolution of this apparent contradiction is that the decay rate of the instanton approach should not be identified with the late-time limit of the instantaneous one, but rather as a weighted average over time, which can significantly differ from the $t \to \infty$ limit at high temperatures.

\section{Two-component atomic condensate}
\label{sec:cond}

It was proposed in~\cite{2015EL....11056001F} that relativistic vacuum decay can be probed experimentally in a two-component Bose-Einstein condensate of cold atoms coupled through a time-dependent electro-magnetic field. 
Under the assumption that density fluctuations are small, and after averaging over the time dependence, the problem reduces to a single real field (the phase difference between the two components) with an emergent Lorentz symmetry at low wave vectors. 
Moreover, the effective potential for this field has two minima, interpreted as a true and a false vacua. 
A possible experimental setup was discussed in~\cite{2015EL....11056001F, 2017JPhB...50b4003F}. 
It was further studied in~\cite{Braden:2017add, Braden:2018tky}, where it was shown to exhibit a parametric instability in a region of parameter space. 

In this section, we show how the parametrized path approach described above can be extended to this model. 
For simplicity, we assume the parameters are such that the instability discussed in~\cite{Braden:2017add, Braden:2018tky} does not occur or has a smaller growth rate than the decay rate $\Gamma$ of the false vacuum. 

\subsection{The model}

One of the difficulties faced when studying vacuum decay is the lack of clear alternatives to the instanton approach, making a test of its results or determination of its domain of validity challenging. 
Ideally, one would have alternative options to determine the decay rate which could be compared with the value obtained from the Euclidean instanton, allowing to quantify its accuracy. 
The approach described above may offer such an alternative. 
Another possibility was proposed in Refs~\cite{2015EL....11056001F, 2017JPhB...50b4003F}. 
As mentioned above, the idea is to simulate relativistic vacuum decay using a two-component Bose-Einstein condensate of cold atoms.
The two atomic states are coupled by a radio-frequency field generating an effective interaction potential. 
For well-chosen parameters, this model exhibits two crucial properties: 
\begin{itemize}
	\item Density fluctuations are small and integrating over them yields a theory on a single real field $\varphi$ with an emergent Lorentz invariance at low frequencies.
	\item The effective potential for $\varphi$ has two non-degenerate minima, interpreted as a false and true vacua.
\end{itemize}
We refer to References~\cite{2015EL....11056001F, 2017JPhB...50b4003F, Braden:2017add} for a more detailed description of the model and a discussion of its physical relevance. 
It was shown in~\cite{Braden:2017add, Braden:2018tky} to have an instability due to parametric resonance between the radio-frequency field and phonic perturbations. 
These references also report an extensive numerical study of the problem using a numerical procedure akin to the truncated Wigner approach commonly used in the Cold Atoms community~\cite{2002JPhB...35.3599S}, showing the interplay between vacuum decay and the parametric resonance. 

In Ref~\cite{Billam:2018pvp}, the calculation of the decay rate is performed in the presence of a vortex. 
It is shown that the Euclidean action of the instanton is systematically smaller than the one obtained without vortex. 
This can be qualitatively understood by noting that the drop in density near the vortex core lowers locally the barrier between the false and true vacua. 
The defect thus catalyses the phase transition, significantly reducing the typical time it takes to occur. 
This is expected to be important for experiments, both because the reduced time-scale helps circumventing stability issues such as the one highlighted in~\cite{Braden:2017add} and because a real experimental system will always have some inhomogeneities, which can act in a qualitatively similar way.
Ref~\cite{Billam:2018pvp} used a simplified model of two-component Bose-Einstein condensate where the effect of the radio-frequency field is averaged and included in an effective potential, and where s-wave interactions between different atomic states are set to zero.
In this section, we will work with the same model. 

The two components are described by complex fields $\psi_{+1}$ and $\psi_{-1}$, sometimes called the condensate wave functions. 
Like wave functions in Quantum Mechanics, they are not observable quantities. 
However, they can be used to define four observables: the atomic densities $\rho_{\pm 1} = \abs{\psi_{\pm 1}}^2$ and local velocities $\mathbf{v}_{\pm 1} = \hbar \s  \nabla \arg (\psi_{\pm 1}) / m$ of each component. 
Conversely, these four quantities can be used to define $\psi_{\pm 1}$ up to two global phases.%
\footnote{In the presence of interactions between the two states, $\arg \lp \psi_{+1}^* \s \psi_{-1} \rp$ may also be observable. 
	There is then only one global $\mathrm{U}(1)$ symmetry.}
The classical Hamiltonian for the system is: 
\begin{equation} \label{eq:4:H}
	H = \int \left[
		\sum_{\ep \in \lb -1, +1 \rb}  \frac{\hbar^2}{2 \s m} \s \big\lvert \nabla \psi_\ep \big\rvert^2
		+ V \lp \psi_{+1}, \psi_{-1} \rp
	\right] \dd^d x,
\end{equation}
where $V$ is an effective potential including the self-interactions between atoms and averaged radio-frequency field.

Reducing the problem to a quantum-mechanical one is technically more involved than the procedure outlined in Section~\ref{sec:Rsf} for a relativistic scalar field because of the need to include density fluctuations. 
Here we only sketch the reduction, and refer the interested reader to Appendix~\ref{app:2c} for the technical details. 
The first step is to write down an action for the system (the Hamiltonian~\eqref{eq:4:H} gives the action of Equation~\eqref{eq:2c:action}). 
We then define the phase difference $\varphi$ between the two condensates, choose an \textit{Ansatz} for $\varphi$ depending on one dynamical parameter $R$, and perform an expansion in $R'$ to compute the densities given $\varphi$. 
The action can then be rewritten in the form~\eqref{eq:Sa}, with functions $K$ and $U$ given respectively by Equations~\eqref{eq:A:K} and~\eqref{eq:A:U}.
One can then follow the procedure outlined in Section~\ref{sec:Rsf} to solve numerically the quantum evolution of $R$, once the potential $V$ and \textit{Ansatz} $\varphi_R$ are chosen. 

Following Ref~\cite{Billam:2018pvp}, we choose a potential of the form
\begin{equation}\label{eq:4:V}
	V \lp \psi_{+1}, \psi_{-1} \rp = 
		\frac{g_0}{2} \s \left[ 
			\lp \abs{\psi_{+1}}^2 - \rho_m \rp^2
			+ \lp \abs{\psi_{-1}}^2 - \rho_m \rp^2
		\right]
		- \la \s \left[ 
			\psi_{+1}^* \s \psi_{-1}
			+ \psi_{-1}^* \s \psi_{+1}
			+ \frac{\eta}{2} \s \lp \psi_{+1} \s \psi_{-1}^* - \psi_{-1}^* \s \psi_{+1}  \rp^2
		\right],
\end{equation}
where $g_0$, $\rho_m$, $\la$, and $\eta$ are positive parameters. 
The first term is the interaction potential for two non-interacting condensates with the same equilibrium density $\rho_m$.
The second term encodes the effect of the radio-frequency field, with an amplitude proportional to $\la$. 
This potential has a true vacuum where the two fields have the same phase and a false vacuum where their phases differ by $\pi$ provided $\la < 2 \s g_0 \s \rho_m$ and $\eta > g_0 / \lp 2 \s \lp g_0 \s \rho_m - \la \rp \rp$. 
The corresponding densities are $\rho_m \pm \la / g_0$, where the $+$ sign corresponds to the true vacuum and the $-$ sign corresponds to the false vacuum.
In the following, for simplicity we work in units where the reduced Planck constant $\hbar$, the atomic mass, and the density $\rho_m$ are equal to $1$.
We work with an \textit{Ansatz} of the form\footnote{Notice that the potential~\eqref{eq:4:V} is invariant under $\varphi \to 2 \s  \pi - \varphi$, and thus under $R \to -R$.}
\begin{equation}
	\varphi_R(r) = \pi \s \left[ 1 - \frac{1}{2} \tanh \lp \frac{r-R}{\sigma} \rp + \frac{1}{2} \tanh \lp \frac{r+R}{\sigma} \rp \right],
\end{equation}
where, as in Section~\ref{sec:Rsf}, $R$ is the dynamical parameter whose evolution is to be determined while $\sigma$ is a static parameter, whose value can be tuned to minimize the Euclidean action.
For $R = 0$, or more generally in the limit $\abs{r / R} \to \infty$, the phase difference is equal to $\pi$. 
Solving for the densities as explained in Appendix~\ref{app:2c} then gives the false vacuum. 
Conversely, in the limit $R \to \pm \infty$ at fixed $r$, $\varphi_R$ goes to $0 \, \mathrm{mod} \, 2 \s \pi$, and solving for the densities gives the true vacuum.

To test the validity of the \textit{Ansatz}, we compared the Euclidean action of the instanton thus obtained to that of the field-theoretical instanton computed in Ref~\cite{Billam:2018pvp}. 
Our results are in agreement within an accuracy of $10\%$ for the range of parameters shown in Figure~3 of that reference. 
Given the exploratory nature of the present analysis and the simplicity of the \textit{Ansatz}, this level of agreement seems satisfactory.

\subsection{Numerical results}

The Schrödinger equation for $R$ can be solved as described in Section~\ref{sec:num}, with the functions $K$ and $U$ replaced by their expressions given in Equations~\eqref{eq:A:K} and~\eqref{eq:A:U}. 
As explained in Appendix~\ref{app:2c}, the main technical difference is that two ordinary differential equations must be solved for each value of $R$ to obtain these functions, inducing some overhead in setting up the initial conditions. 
However, since this calculation needs to be performed only at $t=0$, we find this overhead to be typically small before the total duration of a simulation.

\begin{figure}
	\includegraphics[width=0.49\linewidth]{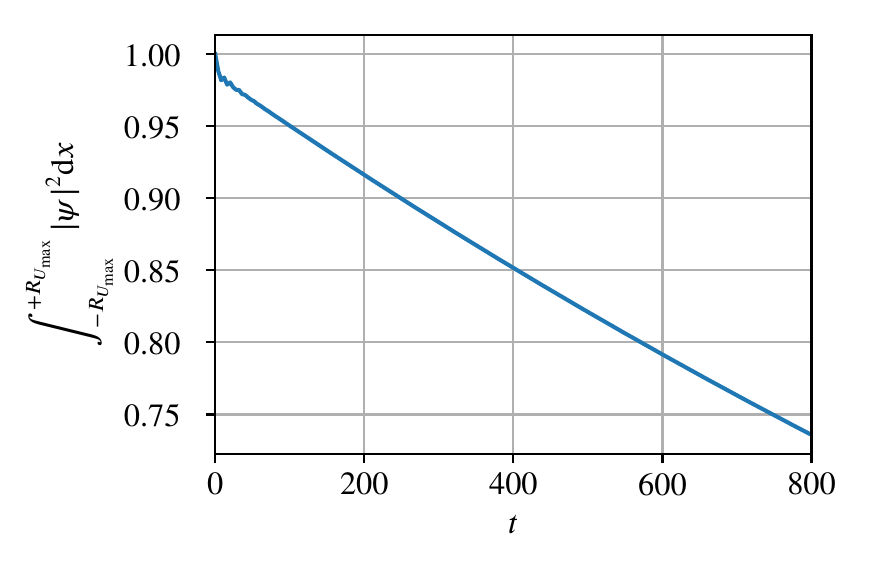}
	\includegraphics[width=0.49\linewidth]{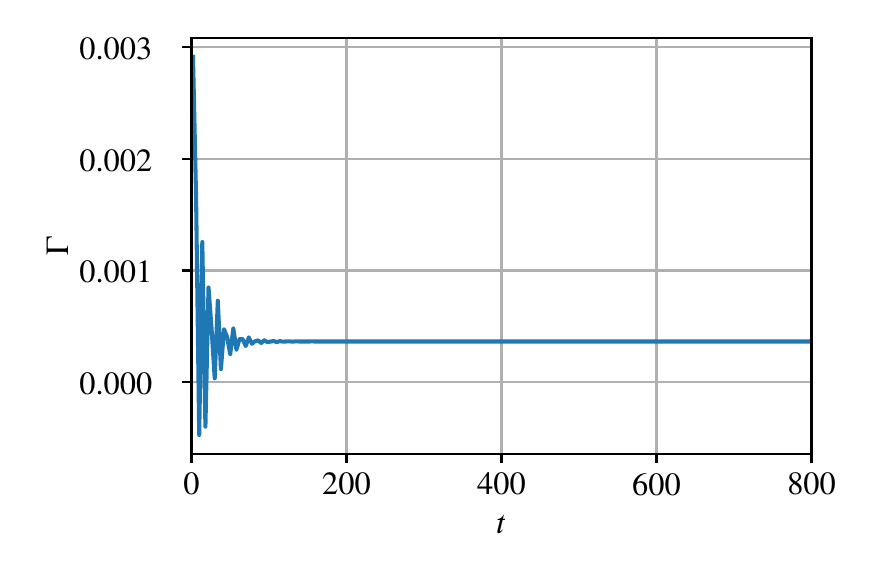}
	\caption{\textit{Left panel:} Probability that $R$ remains between the two values where the maximum of $U$ is reached as a function of time, in a two-component Bose-Einstein condensate with Hamiltonian~\eqref{eq:4:H} and interaction potential~\eqref{eq:4:V}. The parameters are $g_0 = 1$, $\la = 0.25$, and $\eta = 0.6 / (1 - \la)$, in units where $\hbar = m = \rho_m = 1$.  
	The numerical dissipation parameter $\gamma$ is set to $5 \times 10^{-7}$. \\
	\textit{Right panel:} Instantaneous decay rate for the same simulation.}
	\label{fig:pp5b}
\end{figure}

\begin{figure}
	\includegraphics[width=0.49\linewidth]{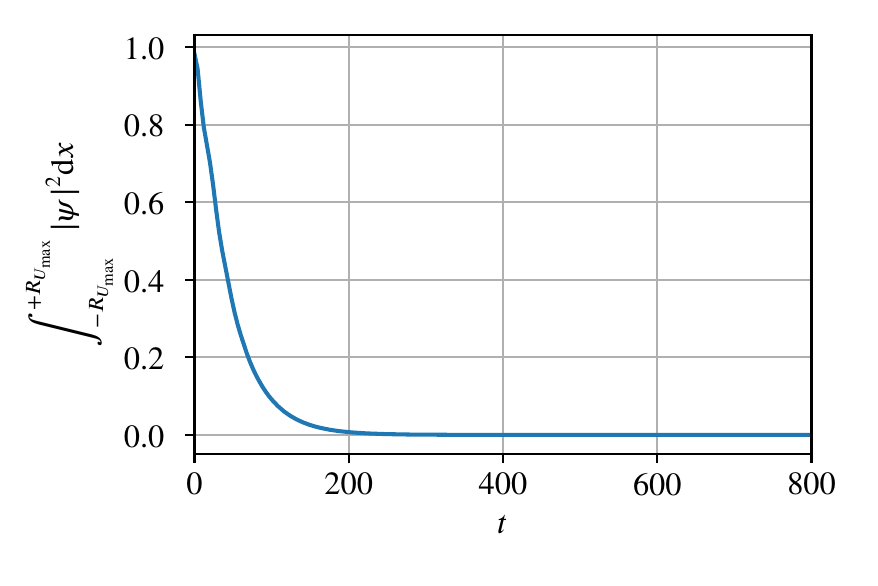}
	\includegraphics[width=0.49\linewidth]{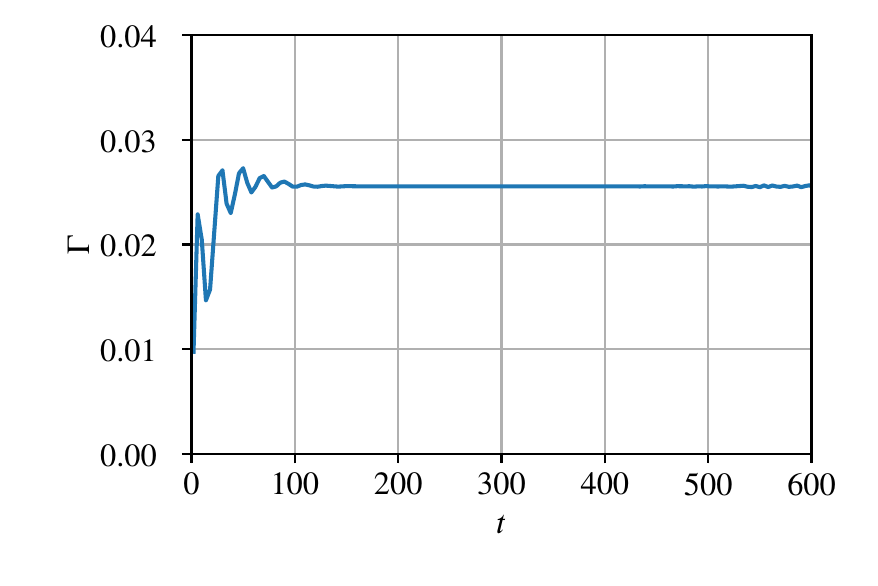}
	\caption{Same plots as in Figure~\ref{fig:pp5b} in the presence of a vortex.}
	\label{fig:pp5c}
\end{figure}

Results for two runs are shown in Figures~\ref{fig:pp5b} and~\ref{fig:pp5c}. 
They are made with the same parameters except for the winding number, equal to $0$ in the first figure (corresponding to a homogeneous false vacuum phase at $t=0$) and $1$ in the second figure (corresponding to a false vacuum with a single vortex; a higher value generally makes the vortex itself unstable).
In both cases, the temperature is equal to $0$. 
Both figures clearly show the dual behaviour already noted in the relativistic case: at relatively early times (here, for $t < 100$), the instantaneous decay rate has large oscillations, while it goes to a constant value, of the same order as that predicted by the instanton approach, at late times. 
The decay rate is significantly larger in the presence of the vortex, confirming the main result of~\cite{Billam:2018pvp}. 
We also note that the vortex seems to tame the early-time oscillations. 
A more systematic study is required to determine if this property is generic, or simply due to the higher saturating value.

\section{Discussion}

In this article, we have developed a simple method to determine the quantum evolution of a scalar field from a false vacuum phase to a true vacuum phase. 
We showed that choosing an \textit{Ansatz} with suitable regularity and integrability properties allows to reduce the field-theory problem to a finite-dimensional quantum-mechanical one, thus greatly simplifying the numerical resolution, and that the obtained decay rate is consistent with the instanton approach. 
We also showed that this procedure can be straightforwardly extended to a cold atom model previously proposed to study vacuum decay experimentally.
The main drawback, both in the relativistic and cold atom cases, is its dependence on the shape of the \textit{Ansatz}: a poor choice is expected to yield smaller decay rates than the real one. 
We argued that this difficulty can be mitigated by making the \textit{Ansatz} dependent on one or several static parameters and optimizing their values. 

Our numerical results indicate that the evolution typically proceeds in two steps. 
At early times, the instantaneous decay rate typically shows large-amplitude oscillations, with an averaged value which, in most cases, is larger than the one predicted by the instanton approach. 
It also strongly depends on the initial state, notably the temperature when working at thermal equilibrium. 
At later times, the instantaneous decay rate tends to a constant value which, in the simulations we performed at zero temperature, seems to be in agreement with the prediction of the instanton approach. 
The compatibility between the two approaches is less straightforward at finite temperature $T$. 
However, we observed that at high $T$ the decay is significantly faster at early times. 
This is at least in qualitative agreement with instanton calculations.

We believe this procedure, which we outlined in two simple models, can be extended to address more challenging questions. 
A first natural extension is to improve the choice of \textit{Ansatz}. 
This can be done in three ways: adding more static parameters to be tuned before running the simulation, more dynamical parameters to be quantized (for instance, one might quantize the parameter $\sigma$ as well as $R$ in the \textit{Ansatz}~\eqref{eq:10:anz3}), or finding an optimal shape through functional calculus techniques. 
In a similar vein, one may consider an \textit{Ansatz} with two or more bubbles of true vacuum to study their interactions. 

A second important extension would be to include gravity. 
Reducing the number of gravitational degrees of freedom to get insight on the quantum evolution was previously proposed in~ \cite{Vilenkin:1982de, Fischler:1989se, Fischler:1990pk, 1990NuPhB.339..417F, Hansen:2009kn}, using for instance a Wheeler-DeWitt equation. 
Including gravitational degrees of freedom into the formalism outlined in Section~\ref{sec:Rsf} would provide a generalisation of these ideas. 
It may, for instance, be used to study bubble nucleation around a black hole, complementing the results of References~\cite{Gregory:2013hja, Burda:2015yfa, Gregory:2016xix, Burda:2016mou, Mukaida:2017bgd, Gregory:2018bdt, Cuspinera:2018woe, Moss:2018rvu, Oshita:2018ptr, 2019arXiv190901378O} by showing the real-time evolution of the quantum state during the bubble nucleation process.

\acknowledgments

I am grateful to Ian Moss, Ruth Gregory, and Renaud Parentani for enlightening discussions and for comments on a preliminary version of this article.
This research was supported by the Leverhulme Trust via the grant RPG-2016-233. 
I thank the Perimeter Institute, where part of this work was done, for its hospitality. Research at Perimeter Institute is supported by the Government of Canada through the Department of Innovation, Science and Economic Development and by the Province of Ontario through the Ministry of Research and Innovation.

\appendix

\section{More numerical results}

\subsection{Results for an asymmetric \textit{Ansatz}}
\label{app:asy}

In the main text, we worked with the symmetric \textit{Ansatz}~\eqref{eq:10:anz3} so that $\phi_R$ goes to the true vacuum in the limits $R \to \pm \infty$. 
To estimate how the precise shape of the \textit{Ansatz} affects the decay rate, in this appendix we show numerical results obtained with the asymmetric\textit{Ansatz} 
\begin{equation} \label{eq:10:anz4}
	\phi_R(r) = \phi_F + \frac{\phi_T - \phi_F}{2} \s \lp 
		\tanh \lp \frac{r + R}{\sigma} \rp 
		- \tanh \lp \frac{r - R}{\sigma} \rp
	\rp.
\end{equation}
It satisfies the conditions of Section~\ref{sub:PP:gen} except the third one, going to the true vacuum as $R \to +\infty$ only. 

\begin{figure}
	\centering
	\includegraphics[width=0.49\linewidth]{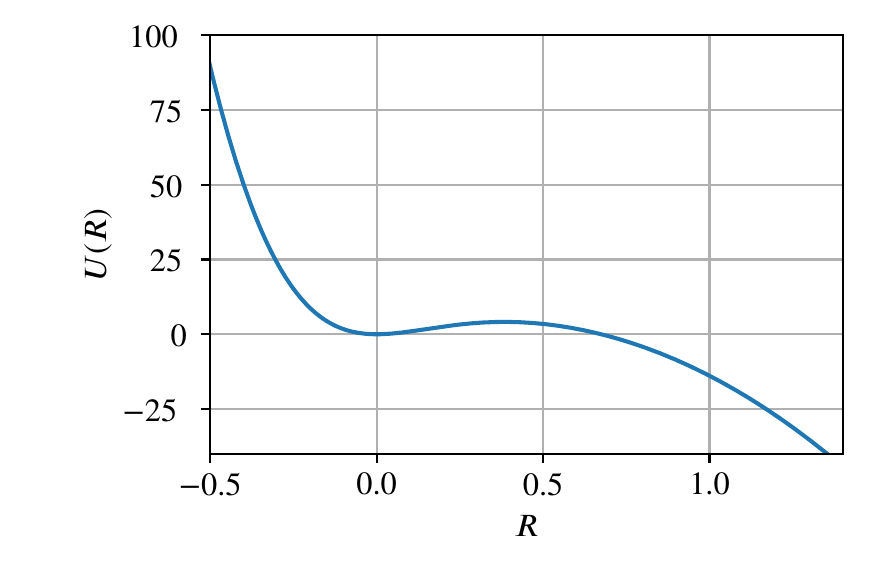}
	\includegraphics[width=0.49\linewidth]{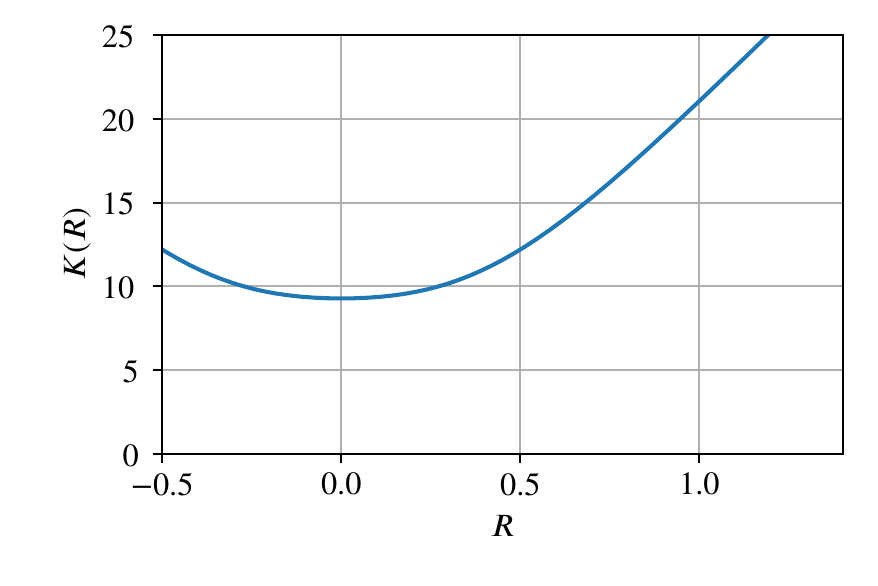}
	\caption{Effective potential $U$ (left panel) and effective mass $K$ (right panel) for the asymmetric \textit{Ansatz}~\eqref{eq:10:anz4} for $\la = 1$, $\eta = 16$, and $\sigma = 0.5$.} 
	\label{fig:10:pp3e_UK}
\end{figure}
The effective potential $U$ and prefactor $K$ of the kinetic term are shown in Figure~\ref{fig:10:pp3e_UK}. 
For $R > 0$, these plots are identical to those in Figure~\ref{fig:UandK}. 
However, for $R < 0$, $U$ is now a monotonically decreasing function.
The function $U$ has only one local maximum, reached at a positive value $R_{U_{\mathrm{max}}}$ of $R$.
We thus change the definition of $P_F$ in Equation~\eqref{eq:PF} to: 
\begin{equation}
	P_F(t) = \int_{-\infty}^{R_{U_{\mathrm{max}}}} \abs{\psi(t,R)}^2 \s \dd R,
\end{equation}
giving the probability that $R$ is smaller than $R_{U_{\mathrm{max}}}$ at time $t$. 
The instantaneous decay rate is still defined by Equation~\eqref{eq:Gamma}.

\begin{figure}
	\centering
	\includegraphics[width=0.49\linewidth]{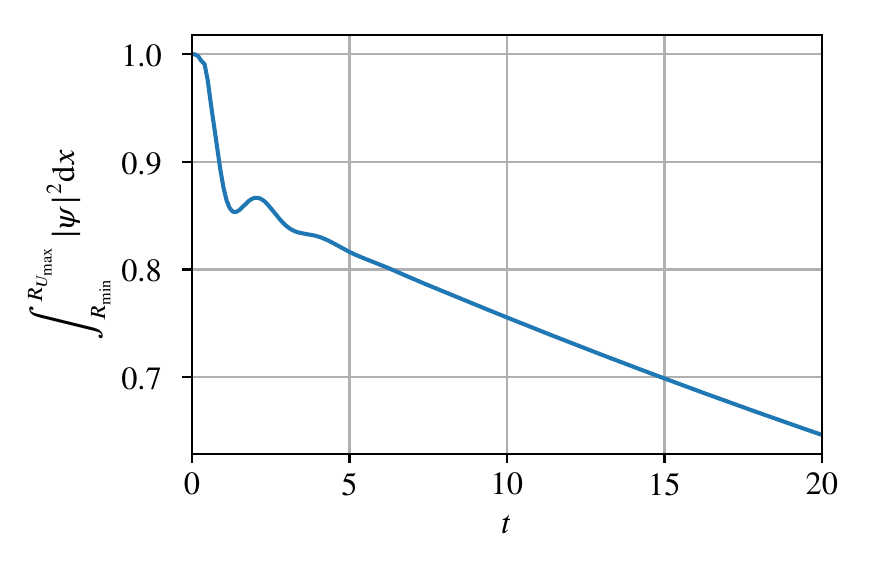}
	\includegraphics[width=0.49\linewidth]{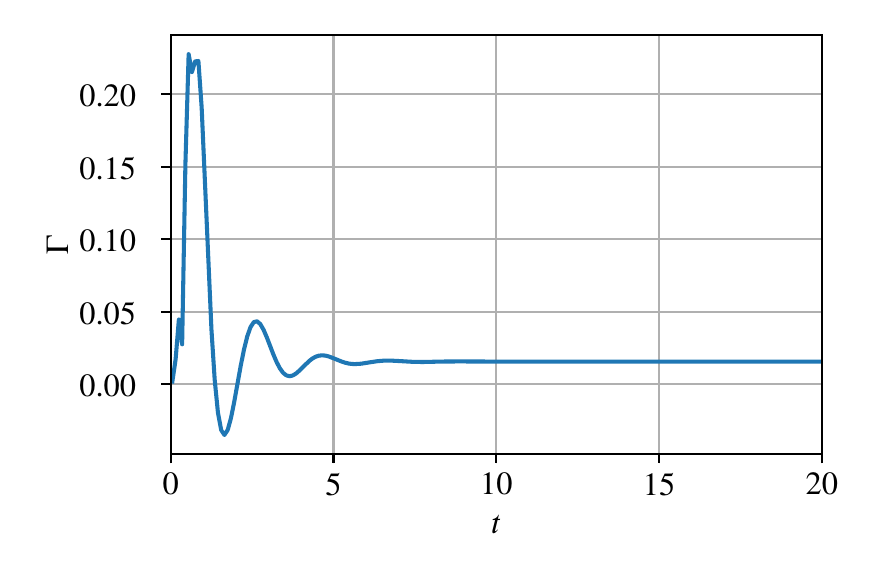}
	\caption{Integral of $\abs{\psi}^2$ up to the point where $U$ reaches its local maximum (left panel) and instantaneous decay rate $\Gamma$ (right panel) for the asymmetric \textit{Ansatz}~\eqref{eq:10:anz4} and the same values of $\la$ and $\eta$ as in Figure~\ref{fig:10:pp3}.
		$R_{\mathrm{min}}$ denotes the lower bound of the numerical grid.}
	\label{fig:10:pp3e_Gamma}
\end{figure}
The probability for $R$ to remain below $R_{U_{\mathrm{max}}}$ and the instantaneous decay rate are shown in Figure~\ref{fig:10:pp3e_Gamma}. 
They both show a relatively good agreement with Figure~\ref{fig:10:pp3}, which indicates that the change of \textit{Ansatz} has little effect on the instantaneous decay rate.

\begin{figure}
	\centering
	\includegraphics[width=0.49\linewidth]{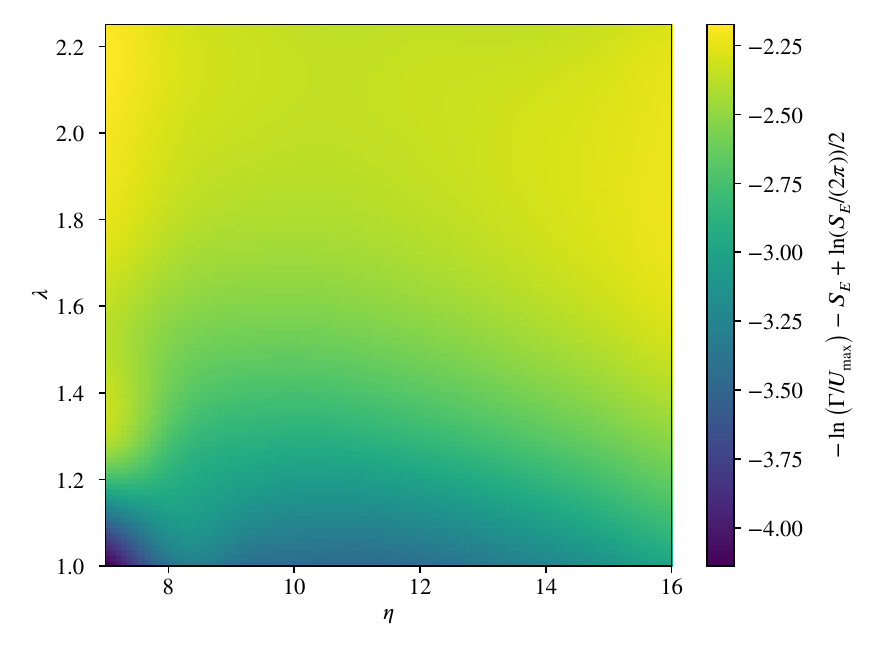}
	\caption{Same plot as in the right panel of Figure~\ref{Fig:10:pp3b} for the asymmetric \textit{Ansatz}~\eqref{eq:10:anz4}.}
	\label{Fig:10:pp3e_ad}
\end{figure}
As done in Figure~\ref{Fig:10:pp3b} for the symmetric \textit{Ansatz}, we show in Figure~\ref{Fig:10:pp3e_ad} the logarithm of the decay rate evaluated at late times, adimensionalized by $U_{\mathrm{max}}$, minus the Euclidean action of the field-theoretic instanton and contribution from one zero mode. 
While the result shows more variability than when using the symmetric \textit{Ansatz}, its range of variation remains well below that of the Euclidean action $S_E$, which varies between $2.0$ and $13.2$. 

\subsection{Results with a radius-dependent width}
\label{app:Rdw} 

\begin{figure}
	\centering
	\includegraphics[width=0.49\linewidth]{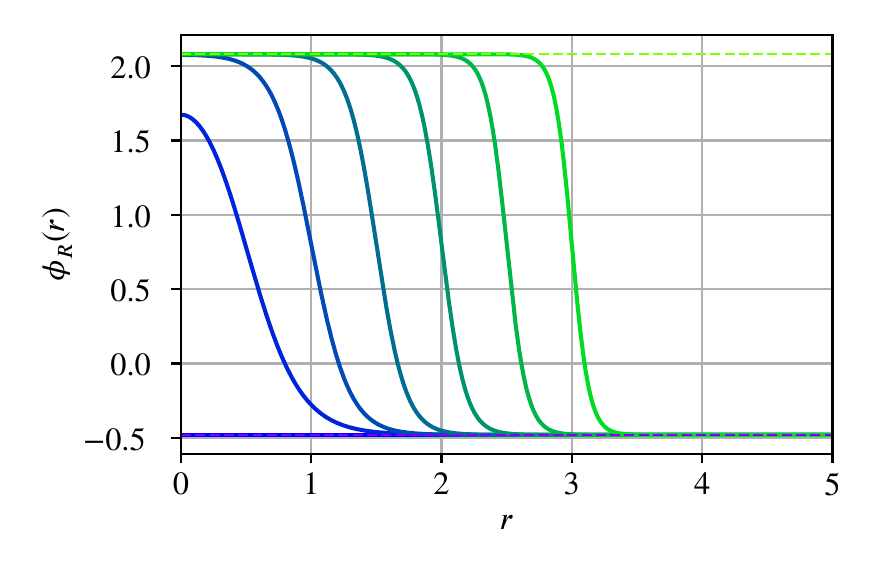}
	\includegraphics[width=0.49\linewidth]{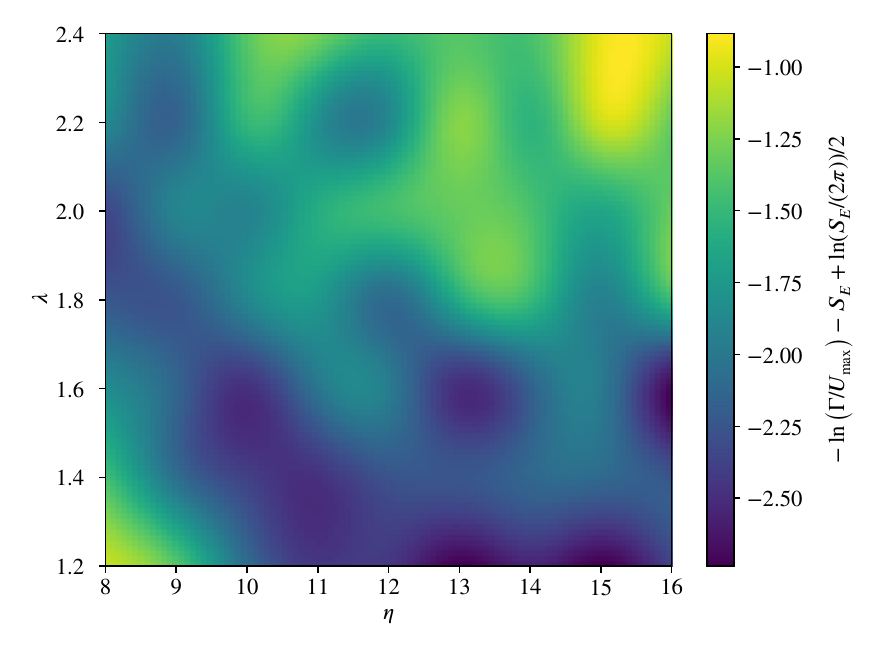}
	\caption{\textit{Left panel:} Ansatz~\eqref{eq:10:anz5} for $\sigma = 0.7$ and different values of $R$. 
		\textit{Right panel:} Same plot as in Figure~\ref{Fig:10:pp3b} for the \textit{Ansatz}~\eqref{eq:10:anz5}. }
	\label{fig:10:pp3f}
\end{figure}

The \textit{Ansatz} used in Section~\ref{sec:num} has a constant width $\sigma$. 
Since relativistic true vacuum bubbles in flat space-time have a width inversely proportional to their radius (this is a consequence of their depending only on the Lorentz-invariant distance from the origin), in this appendix we verify that our numerical results do not significantly change when including such a dependence. 
We work with the \textit{Ansatz}: 
\begin{equation} \label{eq:10:anz5}
	\phi_R(r) = \phi_F + \frac{\phi_T - \phi_F}{2} \s \lp 
		\tanh \lp \frac{(\abs{R}+\sigma) \s (r + \abs{R})}{\sigma^2} \rp 
		- \tanh \lp \frac{(\abs{R}+\sigma) \s (r - \abs{R})}{\sigma^2} \rp
	\rp.
\end{equation}
Like the \textit{Ansatz}~\eqref{eq:10:anz3}, it satisfies all the properties of Section~\ref{sub:PP:gen} except twice differentiability at $R=0$.  
Moreover, the typical width of the bubble wall for $R \gg \sigma$ is $\sigma^2 / R$, showing the same behaviour as that of relativistic bubbles. 
It is shown in the left panel of Figure~\ref{fig:10:pp3f}. 

The numerical calculation of the decay rate proceeds as described in Section~\ref{sec:num}. 
Results are shown in the right panel of Figure~\ref{fig:10:pp3f}. 
They are in relatively good agreement with those of Figure~\ref{Fig:10:pp3b}.

\subsection{Results in four dimensions}
\label{app:3d}

We here show numerical results extending those of Section~\ref{sub:Idr} to $d=3$, \textit{i.e.}, to a 4-dimensional space-time. 
The numerical procedure is the same as that used in the main text for $d=2$. 
In practice, the main two differences are that $U(R)$ goes to $-\infty$ faster in the limit $\abs{R} \to \infty$, in accordance with Equation~\eqref{eq:asyR}, and that decay rates tend to be smaller for given values of $\eta$ and $\la$. 
This makes it challenging to obtain reliable results for small values of these parameters, although in principle possible provided the time and space steps are sufficiently low. 

\begin{figure}
	\centering
	\includegraphics[width=0.49\linewidth]{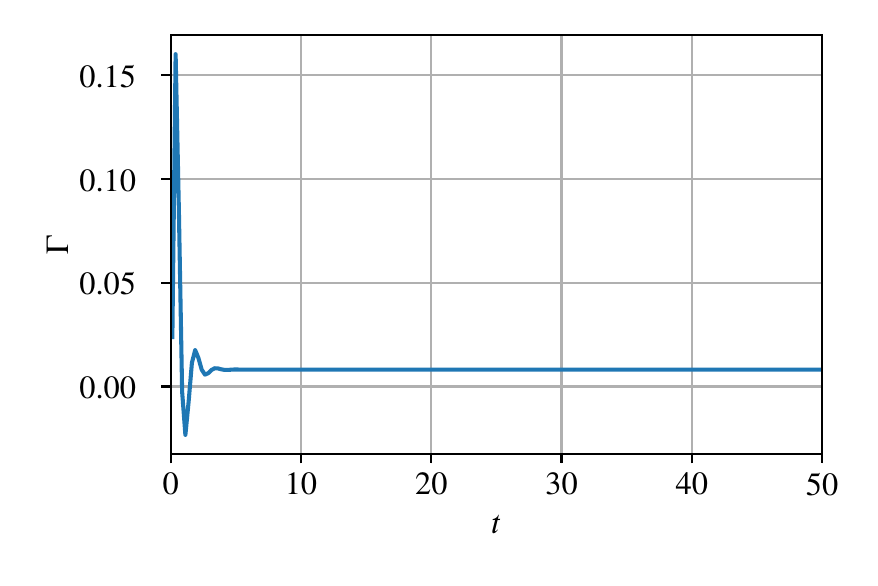}
	\includegraphics[width=0.49\linewidth]{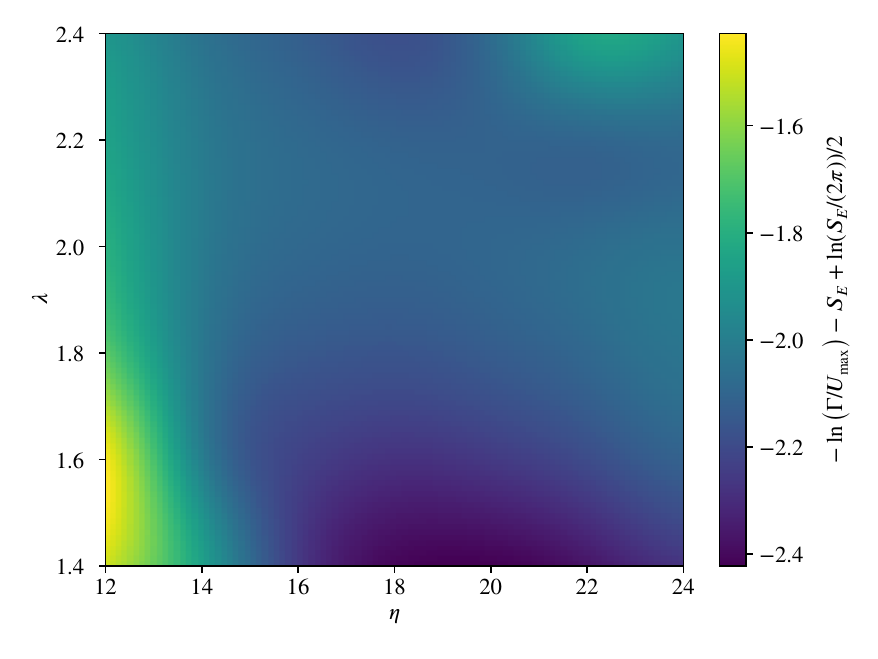}
	\caption{\textit{Left panel:} Plot of the instantaneous decay rate obtained for $d=3$, $\la = 1.5$, and $\eta = 16$ at zero temperature. 
		\textit{Right panel:} Plot of $-\ln \lp \Gamma / U_{\mathrm{max}} \rp - S_E$ plus the contribution from one zero mode, where $\Gamma$ is evaluated at the latest time in each simulation, for $\la$ varying between $1.4$ and $2.4$, and $\eta$ varying between $12$ and $24$.}
	\label{fig:10:pp3c}
\end{figure}

The instantaneous decay rate at zero temperature for $\la = 1.5$ and $\eta = 16$ is shown in the left panel of Figure~\ref{fig:10:pp3c}.
Its behaviour is qualitatively similar to that of the two-dimensional case (see Figure~\ref{fig:10:pp3}), oscillating wildly at early times before reaching a constant value. 
A similar behaviour was observed for the other parameter values we tried. 
The right panel of the Figure compares the opposite logarithm of the asymptotic value of the decay rate, adimensionalized by the maximum of the effective potential $U$, with the quantity
\begin{equation*}
S_E - \frac{1}{2} \ln \lp \frac{S_E}{2 \s \pi} \rp,
\end{equation*}
where $S_E$ is the Euclidean action of the $\mathrm{O}(4)$-symmetric instanton.
(See the discussion of Figure~\ref{Fig:10:pp3b}.) 
Also similarly to the case $d=2$, the difference remains close to $-2$, indicating a good agreement with the instanton approach. 

\section{Parametrized path in the two-component condensate}
\label{app:2c} 

In this appendix we determine the functions $K$ and $U$ in the two-component model of Reference~\cite{Billam:2018pvp}. 
To shorten the expressions, we will drop the dependence of the fields in $r$ and $t$ in intermediate expressions. 
The model is described by two complex fields $\psi_\ep$, $\ep = \pm 1$, with a Lagrangian density
\begin{equation}
	\mathcal{L} = 
		\sum_{\ep \in \lb -1, +1 \rb} \left[ 
			\frac{\ii \s \hbar}{2} \s \lp \psi_\ep^* \s \pd_t \psi_\ep - \psi_\ep \s \pd_t \psi_\ep^* \rp 
			- \frac{\hbar^2}{2 \s m} \s \big\lvert \nabla \psi_\ep \big\rvert^2
		\right]
		- V \lp \big\lvert \psi_{+1} \big\rvert^2, \big\lvert \psi_{-1} \big\rvert^2, \arg(\psi_{+1}) - \arg(\psi_{-1}) \rp,
\end{equation}
where $\hbar$ is the reduced Planck constant, $m$ is the atomic mass, and $V$ is an interaction potential, including the interactions between atoms and the averaged effect of the electro-magnetic field.
We define the densities of the two fields $\rho_\ep = \abs{\psi_\ep}^2$ for $\ep \in \lb -1, +1 \rb$. 
Following~\cite{Billam:2018pvp}, we assume the arguments of $\psi_{\pm 1}$ are equal and opposite up to the contribution of the vortex: there exists a real function $\varphi$ such that $\arg(\psi_{\pm 1}) = \pm \varphi / 2 + n \s \theta$, where $n$ is an integer (equal to $0$ in the absence of vortex) and $\theta$ is an angle coordinate around the vortex core. 
Working in $2$ space dimensions, the action becomes: 
\begin{equation} \label{eq:2c:action}
	\begin{aligned}
		S = - 2 \s \pi \int_{t=-\infty}^{+\infty} \int_{r=0}^\infty & \left[
			\frac{\hbar}{2} \s \lp \rho_{+1} - \rho_{-1} \rp \pd_t \varphi
			+ \frac{\hbar^2}{2 \s m} \s \lp 
				\lp \pd_r \sqrt{\rho_{+1}} \rp^2
				+ \lp \pd_r \sqrt{\rho_{-1}} \rp^2
				+ \lp \rho_{+1} + \rho_{-1} \rp \frac{\lp \pd_r \varphi \rp^2}{4}
			\rp
			\right. \\ & \hspace*{1em} \left. \vphantom{\lp \frac{\lp \pd_r \varphi \rp^2}{4} \rp}
			+ V \lp \rho_{+1}, \rho_{-1}, \varphi \rp
		\right] r \s \dd r \s \dd t.
	\end{aligned}
\end{equation}

Let us now assume we have chosen a series $\varphi_R$ of functions from $\mathbb{R}^+$ to $\mathbb{R}$, indexed by the real variable $R$. 
We consider field configurations where $\varphi$ has the form: 
\begin{equation}
	\varphi (t, r) = \varphi_{R(t)}(r). 
\end{equation}
An additional difficulty compared with the case of Section~\ref{sec:Rsf} is that we have two additional functions $\rho_{\pm 1}$ to determine. 
One could choose an \textit{Ansatz} for these functions too, but this would introduce more arbitrariness. 
Instead, we propose to fix them by extremizing the action at fixed $\varphi$. 
In general, this is complicated by the fact that the Euler-Lagrange equations on $\rho_{\pm 1}$ involve the derivative of $R$, which is not known \textit{a priori}. 
However, as we now show, an expansion in this derivative allows to compute the two densities to leading order using only the value of $R$ at a given time. 

Let $c_0$ be a velocity of the order of the sound speed in the false vacuum phase. 
We expand $\rho_{\pm}$ in powers of $R' / c_0$ in the following way: 
\begin{equation}
	\rho_{\pm 1} = \rho_R^{(\pm 1)} + \gamma_R^{(\pm 1)} \s R' + \chi_R^{(\pm 1)} \s R'^2 + \rho_R^{(\pm 1)} \s O \lp \lp \frac{R'}{c_0} \rp^2 \rp , 
\end{equation}
where, for each value of $R$, $\rho_R^{(\pm 1)}$ are twice differentiable functions from $\mathbb{R}_+$ to $\mathbb{R}_+$ and $\gamma_R^{(\pm 1)}$, $\chi_R^{(\pm 1)}$ are twice differentiable functions from $\mathbb{R}_+$ to $\mathbb{R}$. 
To zeroth order in $R' / c_0$, the action is: 
\begin{equation}
	S^{(0)} = - 2 \s \pi \int_{t = -\infty}^{+\infty} \int_{r=0}^\infty \left[
		\frac{\hbar^2}{2 \s m} \s \lp 
			\lp \pd_r \sqrt{\rho_R^{(+1)}} \rp^2
			+ \lp \pd_r \sqrt{\rho_R^{(-1)}} \rp^2
		\rp
		+ \lp \rho_R^{(+1)} + \rho_R^{(-1)} \rp \frac{\lp \pd_r \Theta_R \rp^2}{4}
		+ V \lp \rho_R^{(+1)}, \rho_R^{(-1)}, \varphi_R \rp
	\right] r \s \dd r \s \dd t. 
\end{equation}
Let us look for field configurations where $\rho_R^{(+1)} = \rho_R^{(-1)}$. 
We call this function $\rho_R$. 
We also assume $\rho_{+1} - \rho_R = \rho_R - \rho_{-1}$, and define $\gamma_R = \gamma_R^{(+1)}$. 
Then, the first order in $R' / c_0$ cancels and the zeroth order becomes: 
\begin{equation}
	S^{(0)} = - 2 \s \pi \int_{t=-\infty}^{+\infty} \int_{r=0}^\infty \left[
		\frac{\hbar^2}{2 \s m} \s \lp 
			\lp \pd_r \sqrt{\rho_R} \rp^2
			+ \frac{\rho_R}{4} \s \lp \pd_r \varphi_R \rp^2
			+ \frac{n^2}{r^2} \s \rho_R
		\rp
		+ V \lp \rho_R, \rho_R, \varphi_R \rp
	\right] r \s \dd r \s \dd t.
\end{equation}
The corresponding Euler-Lagrange equation is: 
\begin{equation}\label{eq:A:rhoR}
	\frac{\hbar^2}{m \s r \s \sqrt{\rho_R(r)}} \s \pd_r \lp r \s \pd_r \sqrt{\rho_R(r)} \rp
	= \frac{\hbar^2}{4 \s m} \s \lp \varphi_R'(r) \rp^2
	+ \frac{\hbar^2 \s n^2}{m \s r^2}
	+ \pd_{\rho_R} V \lp \rho_R(r), \rho_R(r), \varphi_R(r) \rp,
\end{equation}
to be solved with the boundary conditions $\rho_R'(0) = 0$ and $\rho_R(r) \to \rho_F$ as $r \to \infty$, where $\rho_F$ is the density in the false vacuum phase. 

The second-order contribution to the action is: 
\begin{equation}
	\begin{aligned}
		S^{(2)} = {} & - 2 \s \pi \int_{t=-\infty}^{+\infty} \int_{r=0}^\infty \Bigg[
			\hbar \ \gamma_R \s \pd_R \varphi_R
			+ \frac{\hbar^2}{4 \s m} \s \lp 
				\lp \pd_r \lp \frac{\gamma_R}{\sqrt{\rho_R}} \rp \rp^2
				- \lp \pd_r \sqrt{\rho_R} \rp \lp \pd_r \lp \frac{\gamma_R^2}{\rho_R^{3/2}} \rp \rp
			\rp
			 \\ & 
			 + \lp \pd_{\rho_{+1}}^2 V(\rho_R, \rho_R, \varphi_R) - \pd_{\rho_{-1}} \pd_{\rho_{+1}} V(\rho_R, \rho_R, \varphi_R) \rp \gamma_R^2
		\Bigg] \s \lp R' \rp^2 r \s \dd r \s \dd t.
	\end{aligned}
\end{equation}
The corresponding Euler-Lagrange equation is: 
\begin{equation}\label{eq:A:gammaR}
	\begin{aligned}
		& \frac{\hbar^2}{4 \s m \s r \s \sqrt{\rho_R(r)}} \s \pd_r \lp r \s \pd_r \lp \frac{\gamma_R(r)}{\sqrt{\rho_R(r)}} \rp \rp = 
		\frac{\hbar}{2} \s \pd_R \varphi_R(r)
		\\ & \hspace*{1em} {}
		+ \lp 
			\frac{\hbar^2 \s \pd_r \lp r \s \pd_r \sqrt{\rho_R(r)} \rp}{4 \s m \s r \s \rho_R(r)^{3/2}}
			+ \pd_{\rho_{+1}}^2 V (\rho_R(r), \rho_R(r), \varphi_R(r))
			- \pd_{\rho_{-1}} \pd_{\rho_{+1}} V(\rho_R(r), \rho_R(r), \varphi_R(r))
		\rp \gamma_R(r),
	\end{aligned}
\end{equation}
to be solved with the boundary conditions $\gamma_R'(0) = 0$ and $\gamma_R(r) \to 0$ as $r \to \infty$.

Solving Equations~\eqref{eq:A:rhoR} and~\eqref{eq:A:gammaR} with the relevant boundary conditions yields the two functions $\rho_R$ and $\gamma_R$ for each value of $R$. 
We can then define the two functions from $\mathbb{R}$ to $\mathbb{R}$:
\begin{equation}\label{eq:A:K}
	K(R) = - 2 \s \pi \int_0^\infty \hbar \s \gamma_R(r) \s \lp \pd_R \varphi_R(r) \rp r \s \dd r
\end{equation}
and
\begin{equation}\label{eq:A:U}
	U(R) = 2 \s \pi \int_0^\infty \left[ 
		\frac{\hbar^2}{2 \s m} \s \lp
			\lp \pd_r \sqrt{\rho_R(r)} \rp^2
			+ \frac{\rho_R(r)}{4} \s \lp \pd_r \varphi_R(r) \rp^2
		\rp
		+ V(\rho_R(r), \rho_R(r), \varphi_R(r))
		- V(\rho_F, \rho_F, \pi)
	\right] r \s \dd r.
\end{equation}
The difference between the action and that of the homogeneous false vacuum may then be written as 
\begin{equation}
	\Delta S \approx \int_{-\infty}^{+\infty} \left[ 
		\frac{K(R(t))}{2} \s R'(t)^2
		- U(R(t))
	\right] \dd t,
\end{equation}
where terms of order $3$ or more in $R' / c_0$ are discarded.
This expression is similar to Equation~\eqref{eq:Sa}, and we can proceed from here as in Section~\ref{sec:Rsf}. 
The main difference is that we now need to solve two ordinary differential equations for each value of $R$ to obtain $\rho_R$ and $\gamma_R$. 

\bibliography{biblio}

\end{document}